\documentclass[a4paper,onecolumn,11pt,accepted=2017-05-09]{quantumarticle}
\pdfoutput=1
\usepackage[utf8]{inputenc}
\usepackage[english]{babel}
\usepackage[T1]{fontenc}
\usepackage{amsmath}
\usepackage{amsfonts}
\usepackage{hyperref}
\usepackage{graphicx}
\usepackage{subcaption}
\usepackage{caption}
\usepackage{geometry}
\usepackage{float}
\usepackage[title]{appendix}

\usepackage{tikz}
\usepackage{lipsum}
\usepackage{booktabs} 
\usepackage{adjustbox} 
\usepackage{noto}

\begin{document}

\title{Pursuit and Review of Magnetic Resonance Imaging (MRI) based Quantum Computing - Qubit Generation, Spin Purification, Tailored RF Pulses  and  MRI Sequences for Quantum Computing}

\author{Z. H. Cho}
\email{zhcho36@gmail.com}
\homepage{http://quantum-journal.org}
\orcid{0000-0003-0290-4698}
\affiliation{Quantum Computing Research Center, GMRC, Korea University, Seoul, Korea}
\thanks{}

\author{J. H. Han}
\affiliation{Quantum Computing Research Center, GMRC, Korea University, Seoul, Korea}
\orcid{0000-0003-1985-4623}
\author{D. H. Suk}
\affiliation{GMRC, Korea University, Seoul, Korea}
\author{H. J. Jeung1}
\affiliation{Quantum Computing Research Center, GMRC, Korea University, Seoul, Korea}
\author{S. Z. Lee}
\affiliation{Quantum Computing Research Center, GMRC, Korea University, Seoul, Korea}
\orcid{0000-0003-1533-8015}
\author{Y. B. Kim}
\affiliation{School of Medicine, Gachon University, Inchon, Korea}
\author{S. H. Paek}
\affiliation{College of Medicine, Seoul National University, Seoul, Korea}
\author{H. G. Lee}
\affiliation{School of Engineering, Korea University, Seoul, Korea}
\orcid{0000-0002-0335-9508}
\maketitle

\begin{abstract}
We propose a novel MRI (Magnetic Resonance Imaging) technique based quantum bit (qubit) generation with water proton NMR (¹H-NMR), distinct from previously proposed NMR chemical shift or spectroscopic techniques based qubit generation. We briefly review prior NMR-based techniques in the context of quantum computing, focusing on MRI-related methods. The proposed technique utilizes MRI-based gradient methods combined with set of local reverse gradients to generate multiple qubits. This configuration enables the creation of multiple localized constant magnetic fields, each producing a qubit with highly homogeneous field, therefore, a unique single frequency. The RF electronics and signal processing techniques are analogous to those used in conventional MRI scanners, allowing operation at room temperature, with the exception of the main magnet, which remains identical to that of standard MRI systems. Additional advantages of this method include the ability to leverage the extensive pulse techniques and hardware developed for MRI scanners over the past 50 years. Furthermore, the wide array of MRI pulse sequences enables highly sophisticated signal processing for quantum computing applications, such as the spin purification.

The two most widely accepted qubits in the field are the superconducting qubit and the trapped-ion qubit, both of which require either extremely low temperatures with complex instrumentation or a high-vacuum environment to minimize quantum noise. In contrast, the proposed MRI-based qubit generation technique, utilizing water ¹H-NMR, can be implemented using commonly available tools, such as the magnets, gradients, and RF coils found in conventional MRI scanners widely used in the medical community. Additionally, the highest-frequency NMR machine currently available operates at 1.2 GHz (equivalent to 28.0 T) for spectroscopic purposes, suggesting a high signal-to-noise ratio (SNR) for MRI-based qubits. Advances in magnet technology further enhance the potential of MRI-based quantum computing, enabling the generation of signals with higher SNR. This, in turn, facilitates the creation of more precise pseudo-pure or pure-state qubits.

In this paper, we introduce a novel qubit generation scheme and quantum computing (QC) platform based on MRI pulse sequences tailored for quantum computation. These include a modified "Stimulated Echo" technique for spin purification and MRI-compatible pulse sequences for implementing the Hadamard gate to achieve superposition and entanglement—fundamental quantum computing operations. The proposed MRI-based spin purification technique, which employs multiple 90° rotations in a "Stimulated Echo-like" sequence with precise dephasing/rephasing and echo time adjustments, shows promise for generating pseudo-pure to pure spin states that is applicable to quantum computation. This approach is particularly advantageous given the emergence of advanced semiconductor devices, such as 1.5–3 nm semiconductor chips, which enable precise timing control in the proposed MRI-based quantum signal processing. Additionally, we have studied and presented the electronics required for the front-end implementation of MRI-based qubits, providing the tools for practical quantum computation.
\end{abstract}

\section{Introduction}
Historically, modern computers have profoundly transformed human society. Over the past few decades, since the inception of the modern computer in the 1940s, computational power has surpassed human capabilities in many domains, with rapid advancements in both processing speed and memory capacity, now reaching giga- and terabytes and beyond. Computers are integral to daily life, defense, and warfare, influencing human survival. However, certain problems, such as those in molecular sciences or neuroscience, involving 100 billion to several hundred trillion molecules or neurons, exceed the capabilities of modern computers due to their limited computational power. Quantum computing has emerged as a promising solution to address these challenges~\cite{R1,R2,R3,R4,R5,R6,R7}.

Although quantum computing has been studied for over two decades, significant progress has occurred only in recent years, driven by developments such as superconducting qubits based on Josephson junction technology, which operate at extremely low temperatures and high vacuum~\cite{R8,R9,R10,R11,R12,R13}. Another promising approach, the trapped-ion technique, faces limitations in generating large numbers of qubits and requires stringent high-vacuum conditions and precise alignment of qubit atoms in electrostatic and magnetic fields, necessitating further development~\cite{R14}. Other emerging candidates include diamond NV centers, Rydberg atoms, and quantum dots~\cite{R15}.

We propose a novel qubit generation scheme based on magnetic resonance imaging (MRI) techniques, distinct from previously explored NMR spectroscopy-based concepts. This approach, termed ``MRI-Q'' or ``MRQ'' (MRI Quantum Bit), satisfies the requirements for quantum computation~\cite{R16}. A key advantage of MRQ is the ability to generate large numbers of qubits that can operate at room temperature, eliminating the need for specialized low-temperature environments required by superconducting or trapped-ion qubits~\cite{R8,R9,R10,R11,R12,R13,R14,R15}.

Despite its potential, MRI-based qubit generation faces skepticism due to previous unsuccessful attempts to develop NMR-based qubits for quantum computing. These efforts, primarily using NMR spectroscopic or chemical shift techniques (e.g., \textsuperscript{13}C-NMR with trichloroethane (C\textsubscript{2}HCl\textsubscript{3}) or (2,3)-dibromothiophene), see Fig1(a), failed to produce sufficient number of qubits or achieve ``pure'' state qubits suitable for quantum computations~\cite{R17,R18,R19,R20,R21}. In contrast, our proposed MRI-based approach, illustrated in Fig.1(b), leverages MRI gradient techniques to generate multiple qubits~\cite{R16}.

A significant challenge in NMR- or MRI-based quantum computing is achieving highly ``pure'' state qubits, as NMR relies on ensembles of nuclear spins with spin-$\frac{1}{2}$ properties that exhibit statistical behavior under strong magnetic fields. This ensemble nature has fueled skepticism in the quantum computing community regarding NMR-based qubits, particularly following the success of superconducting qubits, such as the transmon used by IBM and Google~\cite{R8,R9,R10,R11,R12,R13,R17,R18,R19,R20,R21}. The difficulty of generating pure state qubits with NMR or MRI stems from their reliance on bulk or ensembles nature of spins rather than single atom ~\cite{R14,R15}.

In the following sections, we review key challenges and considerations in quantum computation.

\subsection{Spin Distributions and Spins with “Pseudo Pure”  or  “Pure” States.}

In quantum computation, we customarily define  “Pseudo-Pure” and “Pure” states of the qubit which often can be defined by the Density Matrix r (17,18,19) , i.e.,

\begin{equation}
\rho = \frac{1 - \varepsilon}{2^N} + \varepsilon (\left| \psi \right\rangle \left\langle \psi \right|)
\end{equation}

In quantum computing, the density matrix describes an ensemble of nuclear spins, where \( N \) is the total number of spins, and \( \epsilon \) represents the fraction of spins in a ``pure'' state. The wave functions \( \langle \Psi | \) and \( | \Psi \rangle \) denote the bra and ket of pure states, respectively. Typically, the density matrix characterizes a system where most molecules are in statistically distributed states, with only a small fraction, \( \epsilon \), is in a pure state. For quantum computation, this ``pure'' state is generally required, as it enables reliable computation. The unitary transformation of the density matrix, as shown in Eq.2, facilitates quantum operations essential for computation.

\begin{equation}
U(t)\rho U(t)^{-1} = \frac{1 - \varepsilon}{2^N} + \varepsilon (\left| U(t)\Psi \right\rangle \left\langle U(t)^{-1} \Psi \right|)
\end{equation}

The state described by Eq.2 is known as the ``pseudo-pure'' state, however, it often referred to as a ``pure'' state, since the wave functions \( |\Psi\rangle \) and \( \langle\Psi| \) represent pure states. This suggests that unitary operations can transform the system into a pure state, making it suitable for quantum computation. Among the unitary operators, rotation operators---such as multiple $90^\circ$ rotations or Stimulated Echo pulse sequences used in MRI---are particularly relevant for quantum computing applications~\cite{R27,R28,R29}.

In this paper, we discuss recent progress in quantum qubit development based on magnetic resonance imaging (MRI) techniques, distinct from traditional NMR spectroscopy, using water protons as the source for quantum bits (qubits). We focus on the MRI Quantum Bit (MRQ), emphasizing qubit generation and unitary operators applicable to quantum computation, such as those enabling superposition and entanglement---central components of quantum computing~\cite{R19,R20,R21}. Fig.2(a) illustrates a preliminary MRI-based qubit generation scheme, previously proposed, which uses a magnet, a gradient, and water bars as the qubit source~\cite{R16}. Fig.2(b) presents an advanced configuration incorporating additional local reverse gradients to compensate for local field inhomogeneities introduced by the main gradient. This setup enables the creation of multiple homogeneous and constant local fields, each of which can produce a qubit with a unique single frequency~\cite{R16,R26,R27}. Note the local reverse gradients $G_1(z_1), G_2(z_2),...$ on the original local gradients $G_1(z_1), G_2(z_2),...$(See Fig 2(b)).

Fig.3 depicts the overall quantum computing platform, incorporating multiple qubits generated using the main gradient, local reverse gradients (as shown in Fig.2(b)), and RF coils for MR signal control and reception. Small local proton bars, positioned within the Q-coils, serve as qubit sources. These local constant fields, or ``qubit areas,'' are equipped with small RF coils and a large main RF coil. This configuration allows the generation of multiple qubits, each with distinct frequencies (e.g., \( \omega_1, \omega_2, \ldots \)), corresponding to unique local constant fields (e.g., \( B_1(z_1), B_2(z_2), \ldots \))~\cite{R16,R26,R27}.

\subsection{Spin Distributions and Status of “Bulk”, “Pseudo Pure”  and “Pure” States of Spins. }

The spin distribution states in the absence of a magnetic field (\( B = 0 \)) and in the presence of a magnetic field (\( B = B_0 \)) are illustrated in Fig.4(a) and Fig.4(b), respectively. At \( B = 0 \), a drop of water containing \( N = 2 \times 3.35 \times 10^{22} \) spins exhibits a random distribution, as shown in Fig.4a). In contrast, at \( B = B_0 \), the spins align directionally, as depicted in Fig.4(b), with the majority oriented toward either the low-energy state (upward, yellow) or the high-energy state (downward, red). This polarized configuration, known as Zeeman splitting, results in spin populations \( n_{\text{OL}} \) (low-energy) and \( n_{\text{OH}} \) (high-energy), where \( n_{\text{OL}} > n_{\text{OH}} \).

Fig.5 provides an expanded view of the low-energy spin population \( n_{\text{OL}} \), highlighting its \( x \)- and \( y \)-directional components, \( n_{\text{Ox}} \) and \( n_{\text{Oy}} \), which are used for MR signal processing. These spins, referred to as ``bulk spins,'' are distinct from the total spin population \( N \), with the relation \( N \gg n_{\text{OL}} + n_{\text{OH}} \). By applying unitary operators, such as $90^\circ$ \( x \)- or \( y \)-directional RF pulses, to \( n_{\text{OL}} \), we can isolate specific spin groups, \( n_{1x} \) and \( n_{1y} \), as shown in Fig.6. These groups, termed ``Pseudo-Pure-1'' spins, are highly selected \( x \)- or \( y \)-component spin populations suitable for MRI-based quantum signal processing.

\subsection{Spin Purification with Unitary Operators :  $90^\circ$ Rotation and Modified Stimulated Echo Techniques .}

The spin purification process at the ``bulk'' spin level, illustrated in Fig.6, achieves relatively coarse spin selection but falls short of the high-level purification required for quantum computation. Fig.7 depicts a high-level spin purification technique using a unitary operator, specifically a $90^\circ$ \( x \)-rotation. Initially, a part of bulk spins with a first RF pulse, \( R_x(90) \) (denoted \( a_0 \)), followed by a dephasing period \( t_{a1} \). This dephasing enables the selection of a highly purified spin pair \( a-a' \) aligned along the \( x \)-axis. A subsequent RF pulse \( a_1 \) rotates all spins in the \( x \)-\( y \) plane by $90^\circ$ at time \( t = t_1 \), except for the \( a-a' \) spin pair, effectively isolates the spin pair. After an additional waiting period \( t_{a2} \), equal to \( t_{a1} \), the refocused \( a-a' \) spin pair is obtained, as shown on the right of Fig.7.

Multiple repetitions of this process lead to progressively purified spin pair selection, as illustrated in Fig.8, where three consecutive $90^\circ$ \( x \)-rotation RF pulses, combined with two $180^\circ$ RF pulses, form a ``Modified Stimulated Echo'' sequence~\cite{R29}. In Fig.8, First, three $90^\circ$ RF pulses with two $180^\circ$ RF pulses are applied with precisely adjusted time intervals (\( t_{a1}, t_{a2}, t_{a3}, t_{a4} \)) for dephasing and rephasing. For simplicity, we assume a purification factor of \( 10^6 \) at each step. A critical aspect of this technique is the precise control of spin pair selection, requiring dephasing and rephasing times in the nano- to picosecond range in an NMR or MRI environment.

To achieve highly purified spins, carefully timed rotation pulses in a noiseless environment using, such as the ``Stimulated Echo'' technique will be essential. As seen, for completeness, to achieve an enhanced purification method we have incorporated two additional $180^\circ$ spin echoes to correct errors during the dephasing and rephasing periods, as shown in Fig.8 (note the additional $180^\circ$ RF pulses, \( a_2 \) and \( a_4 \), for spin echoes). In this ``Modified Stimulated Echo'' technique, the dephasing periods \( t_{a1} \) and \( t_{a3} \) are set equal to ensure compensation of errors through opposing spin echoes. Advances in semiconductor devices, such as modern nano-chips, enable us to control precisely in the picosecond range. Note that the extent of achievable spin purity and its sufficiency for error-free quantum computation---enabling superposition, entanglement, and error correction---remains a critical question for the future quantum computing.

\subsection{Realization of Quantum Superposition with Hadamard Gates and Related MRI Pulse Sequences -  Unitary Operators in MRI  : Rotation Operators , \(Rx(\pi ), Ry(\pi/2)\), and Hadamard Operators.}

One of the most challenging aspects of quantum computation in an MRI environment is achieving superposition via the Hadamard gate. Fig.9(a) and Fig.9(b) illustrate two exemplary applications of the Hadamard gate, one for \( H|0\rangle \) and the other for \( H|1\rangle \). These figures demonstrate how rotation gates, specifically \( R_x(\pi) \) and \( R_y(\pi/2) \), are used to implement the Hadamard operation, which is mathematically expressed as:

\begin{equation}
H = i R_x(\pi) R_y\left(\frac{\pi}{2}\right)
= (i)(-i)
\begin{bmatrix}
0 & 1 \\
1 & 0
\end{bmatrix}
\cdot \frac{1}{\sqrt{2}}
\begin{bmatrix}
1 & -1 \\
1 & 1
\end{bmatrix}
\end{equation}

For |0>,

\begin{equation}
\begin{aligned}
H|0>
&= i R_x(\pi) R_y\left(\frac{\pi}{2}\right) |0> \\
&= (i)(-i)
\begin{bmatrix}
0 & 1 \\
1 & 0
\end{bmatrix}
\cdot \frac{1}{\sqrt{2}}
\begin{bmatrix}
1 & -1 \\
1 & 1
\end{bmatrix}
\begin{bmatrix}
1 \\
0
\end{bmatrix} \\
&= \frac{1}{\sqrt{2}}
\begin{bmatrix}
1 & 1 \\
1 & -1
\end{bmatrix}
\begin{bmatrix}
1 \\
0
\end{bmatrix}
= \frac{1}{\sqrt{2}}
\begin{bmatrix}
1 \\
1
\end{bmatrix}
= |\Psi_{+x}>
\end{aligned}
\end{equation}

Corresponding spin trajectory of this Hadamard operation for |0> is given in Fig. 9(a). While operation for |1> is given in the following and its spin trajectory is shown in Fig.9(b).

\begin{equation}
\begin{aligned}
H|1>
&= i R_x(\pi) R_y\left(\frac{\pi}{2}\right) |1> \\
&= (i)(-i)
\begin{bmatrix}
0 & 1 \\
1 & 0
\end{bmatrix}
\cdot \frac{1}{\sqrt{2}}
\begin{bmatrix}
1 & -1 \\
1 & 1
\end{bmatrix}
\begin{bmatrix}
0 \\
1
\end{bmatrix} \\
&= \frac{1}{\sqrt{2}}
\begin{bmatrix}
1 & 1 \\
1 & -1
\end{bmatrix}
\begin{bmatrix}
0 \\
1
\end{bmatrix}
= \frac{1}{\sqrt{2}}
\begin{bmatrix}
1 \\
-1
\end{bmatrix}
= |\Psi_{-x}>
\end{aligned}
\end{equation}

As seen, we have successfully realized \( |\Psi_{-x}\!> \) for \( |1\!>\).

\subsection{Another Approach to Hadamard with Utilization of MR Specific Property, the T1 Recovery and T2 Relaxations}

In MRI, a Hadamard-like operation can be achieved using a distinct sequence that leverages the \( T_1 \) recovery process, as illustrated in Fig.10. For instance, an \( R_y(90) \) pulse is applied, followed by a delay time of \( t = 0.693 T_1 \), which corresponds to the time when the longitudinal magnetization recovers to half its maximum value, i.e., \( S = 0.5 \), as shown in Fig.10(b). The spin trajectories and pulse sequences are depicted in Fig.10(a) and Fig.10(b), illustrating the spin positions at \( t = 0.693 T_1 \), denoted as \( t_{1-} \) (or \( t_{1+} \)). At this point, the \( x \)-component of the spins, \( n_{0x} \), is rotated by $90^\circ$\(_y\) to align with the \( y \)-component, \( n_{1y} \), forming a Hadamard operation, as shown on the right side of Fig.10(b). This approach yields the same result as the Hadamard operation depicted in Fig.9(a).

\section{Technical Descriptions}
\subsection{MRI based Quantum Bit (Qubit) Generation Scheme and Computing Platform.}

As described earlier, the basic quantum bit (qubit) generation schemes using magnetic resonance imaging (MRI) techniques, recently proposed, are illustrated in Fig.2 and Fig.3 (see also the Appendix). The overall scheme of the MRI-based quantum computing platform, incorporating a set of gradient coils (main and local reverse gradient coils) and RF coils (main and Q-coils), is shown in Fig.3\cite{R16,R22,R23,R24,R25,R26,R27,R28}. A key innovation is the gradient configuration, which enables the generation of multiple qubits within MRI-like magnetic environments. These environments consist of a main magnetic field \( B_0 \), a main gradient, and a set of local reverse gradients, producing small, magnetically homogeneous local fields, \( B_1(z_1), B_2(z_2), \ldots \), along the \( z \)-direction (extendable to \( x \)- and \( y \)-directions)~\cite{R16}. Each homogeneous field region can host a qubit, paired with corresponding small RF coils (e.g., \( \text{RF}_1, \text{RF}_2 \), or Q-coils) to form multiple qubits. Additionally, small water bars (local proton bars) are placed at the center of each Q-coil. This configuration, combined with appropriate excitation and receiving RF coils (Q-coils), enables the generation of multiple qubits necessary for quantum computation in an MRI-like environment, analogous to a conventional MRI scanner~\cite{R25,R26,R27,R28}. As shown in Fig.2 and Fig.3, the local magnetic fields \( B_1(z_1), B_2(z_2), \ldots \), created by the main magnet (\( B_0 \)), main gradients (\( G(z_1), G(z_2), \ldots \)), and local reverse gradients (\( G_1(z_1), G_2(z_2), \ldots \)), can be expressed mathematically as,

\begin{equation}
\left.
\begin{aligned}
B_1(z_1) &= B_0 + \Delta B_1(z_1) = B_0 + G(z_1) - G_1(z_1) \\
B_2(z_2) &= B_0 + \Delta B_2(z_2) = B_0 + G(z_2) - G_1(z_2) \\
B_3(z_3) &= B_0 + \Delta B_3(z_3) = B_0 + G(z_3) - G_3(z_3) \\[0.5em]
&\vdots \\[0.5em]
B_n(z_n) &= B_0 + \Delta B_n(z_n) = B_0 + G(z_n) - G_n(z_n)
\end{aligned}
\right\}
\end{equation}

where \(\Delta B_1(z_1)\),\(\Delta B_2(z_2)\),... etc are the discrete constant homogeneous field created by main gradient \(G(z)\) and reverse gradient \(G_1(z_1)\) etc.
These local magnetic fields \( B_1(z_1), B_2(z_2), \ldots \), are created by the main magnet (\( B_0 \)), main gradients (\( G(z_1), G(z_2), \ldots \)), and local reverse gradients (\( G_1(z_1), G_2(z_2), \ldots \)). These local magnetic fields generate multiple qubits with corresponding Larmor frequencies \( \omega_1, \omega_2, \ldots \), given by:

\begin{equation}
\left.
\begin{aligned}
\omega_1 = \gamma B_1(z_1) \\
\omega_2 = \gamma B_2(z_2) \\
\omega_3 = \gamma B_3(z_3) \\[0.5em]
&\vdots \\[0.5em]
\omega_n = \gamma B_n(z_n)
\end{aligned}
\right\}
\end{equation}

where \(\gamma\) is the gyromagnetic ratio.

\subsection{MR Qubit Input – Output Couplings and Electronic Circuits for the Implementation of Quantum Computing}

To implement the MRI or NMR based Quantum Computation, properly coupling the MRQ or qubit to the existing electronics and circuitry is important, especially in the front-end such as control inputs and qubit signal readouts. In Fig. 11, the front-end coupling of a qubit is illustrated. MR Qubit will be controlled by x or y inputs such as $90^\circ_y$ or $180^\circ_x$ RF pulses and qubit signals or outputs are readout via an additional readout RF coil located in perpendicular to z-axis, the x-y plane, to maximize the signal output (see Fig. 11). Within this simplified illustration we have shown the locations and directions of main magnetic field B0 , main gradient G(z), local reverse gradients  G1(z1), RF coils (x and  y) and receiver coil, and the proton source H2O.  One MR Qubit, therefore, consists of, (see table1),

\begin{table}[h]
\centering
\small 
\caption{Components of One MR Qubit}
\label{tab:mr_qubit_components}
\begin{adjustbox}{max width=\textwidth} 
\begin{tabular}{p{2.5cm}p{2.5cm}p{3cm}p{3cm}p{2.5cm}} 
\toprule
\textbf{Qubit-Signal Source} & \textbf{Qubit Magnet} & \textbf{Qubit Gradients} & \textbf{Control Inputs} & \textbf{Qubit Output} \\
\midrule
A drop of water (H\textsubscript{2}O) & A magnet with field \( B_0 \) & A main gradient and a local gradient & A set of \( x \) and \( y \)-RF coil & A receiver RF coil \\
\bottomrule
\end{tabular}
\end{adjustbox}
\end{table}

In Fig. 12, another simplified illustration of the qubit and its couplings to form a basic operation, the Hadamard gate, is  given, as an example. (see also Appendix).

\subsection{Transformation of Qubit Outputs and Control and Target Signals for the Applications to MRI based Quantum Signal Processing.}

To implement quantum computation in an MRI setting, it is essential to transform qubit signals into MRI-compatible pulses that enable quantum operations. Qubit signals, such as \( |0\rangle \) and \( |1\rangle \), which represent the cosine and sine components, are converted into ``double'' or ``single'' RF pulses of $180^\circ$ or $90^\circ$, respectively, to align with MRI pulse sequences, as shown in Fig.13. It is also critical to distinguish whether pulses belong to the ``control'' or ``target'' parts, as illustrated in Fig.13. For the control part, cosine and sine signals are transformed into double or single RF pulses of $180^\circ$ or $90^\circ$, respectively. For the target part, however, pulses are configured as a quadrupole form, comprising two $90^\circ$ pulses and two $180^\circ$ pulses, as depicted in the lower part of Fig.13. A summary of the cosine and sine pulses for the control and target parts are given as.

\begin{equation}
\text{Control: }
\begin{aligned}
\cos(\varpi t)\ \text{or}\ |0\rangle &\rightarrow R(180) + R(180) \\
\end{aligned}
\end{equation}

\begin{equation}
\text{Control: }
\begin{aligned}
\sin(\varpi t)\ \text{or}\ |1\rangle &\rightarrow R(90)\ \text{or}\ -R(90)
\end{aligned}
\end{equation}

\begin{equation}
\text{Target: }
\begin{aligned}
\cos(\varpi t)\ \text{or}\ |0\rangle &\rightarrow R(90) + R(180) + R(180) + R(-90) \\
\end{aligned}
\end{equation}

\begin{equation}
\text{Target: }
\begin{aligned}
\sin(\varpi t)\ \text{or}\ |1\rangle &\rightarrow R(90) + R(180) + R(180) + R(-90)
\end{aligned}
\end{equation}

These representative MRI compatible transformed  RF pulses will be used throughout the quantum computations such as “Entanglement” which will be discussed in the followings.

\subsection{Hadamard and CNOT Operations with Transformed Output Pulses of the Control and Target Signals for the Bell State formation}

In Fig, 14 (a), an example of quantum computation using these transformed pulses for the implementation of “Superposition” and “Entanglement”  via Hadamard and CNOT operations is shown. This diagram illustrates an example of two inputs, one for “Control \( |\phi_1> \)” and the other for ”Target \( |\phi_2> \)” each with                          |0> + |1>. As it is known, this exemplary case is the well known “Bell states” with which quantum computation can be performed. As shown, in the right side of the figure, the resulting four Bell states are shown with combinatorial multiplication steps for the ‘Entanglement”  output, the Bell states that are noted as ; \( |\Phi^+\rangle,|\Phi^-\rangle,|\Psi^+\rangle,|\Psi^-\rangle\).  In Fig. 14(b), a summary of the numerical computational steps are shown for reference.

First, in the Control side, the Hadamard outputs are given,
\begin{equation}
H|0> = \left[ B+ + B{\overset{+}{+}} \right], \quad B+ = |0>, \quad B{\overset{+}{+}} = |1>
\end{equation}

\begin{equation}
H|1> = \left[ B- + B{\overset{-}{-}} \right], \quad B- = |0>, \quad B{\overset{-}{-}} = -|1>
\end{equation}

From Eqs. 12 and 13 , we find 4 outputs which will function as “Control Signals”, that is,

\begin{equation}
B+,  B-,  B{\overset{+}{+}}, B{\overset{-}{-}}
\end{equation}

In the Target side, we also find,

\begin{equation}
\begin{aligned}
I|0\rangle &= [A+] \quad : \quad A+ = |0\rangle \\
I|0\rangle &= [A-] \quad : \quad A- = |1\rangle
\end{aligned}
\end{equation}

From Eq. 15, we find 2 outputs which will function as “Target Signals”, that is,

\begin{equation}
A+, A-
\end{equation}

Together with CNOT gates, 4-Control outputs and 2-Target outputs will form  8- CNOT outputs which are given as,

\begin{equation}
\begin{aligned}
&C[B+A+],\; C[B-A+],\; C[B+A-],\; C[B-A-], C[B{\overset{+}{+}}A+],\; C[B{\overset{-}{-}}A+],\; C[B{\overset{+}{+}}A-],\; C[B{\overset{-}{-}}A-]
\end{aligned}
\end{equation}

Combination of the two, Eqs. 16 and 17 , we find 8 outputs which will form 4-Bell states, that is,

\[
\begin{aligned}
&C[B+A+],\; C[B-A+],\; C[B+A-],\; C[B-A-], \;C[B{\overset{+}{+}}A+],\; C[B{\overset{-}{-}}A+],\; C[B{\overset{+}{+}}A-],\; C[B{\overset{-}{-}}A-]
\end{aligned}
\]
\begin{equation}
|\Phi^+\rangle = |\Phi^+\rangle^{0} + |\Phi^+\rangle^{1}
\end{equation}

\[
\begin{aligned}
&C[B+A+],\; C[B-A+],\; C[B+A-],\; C[B-A-], \; C[B{\overset{+}{+}}A+],\; C[B{\overset{-}{-}}A+],\; C[B{\overset{+}{+}}A-],\; C[B{\overset{-}{-}}A-]
\end{aligned}
\]
\begin{equation}
||\Phi^-\rangle = |\Phi^-\rangle^{0} + |\Phi^-\rangle^{1}
\end{equation}

\[
\begin{aligned}
&C[B+A+],\; C[B-A+],\; C[B+A-],\; C[B-A-],\; C[B{\overset{+}{+}}A+],\; C[B{\overset{-}{-}}A+],\; C[B{\overset{+}{+}}A-],\; C[B{\overset{-}{-}}A-]
\end{aligned}
\]
\begin{equation}
|\Psi^+\rangle = |\Psi^+\rangle^{0} + |\Psi^+\rangle^{1}
\end{equation}

\[
\begin{aligned}
&C[B+A+],\; C[B-A+],\; C[B+A-],\; C[B-A-],\; C[B{\overset{+}{+}}A+],\; C[B{\overset{-}{-}}A+],\; C[B{\overset{+}{+}}A-],\; C[B{\overset{-}{-}}A-]
\end{aligned}
\]
\begin{equation}
|\Psi^-\rangle = |\Psi^-\rangle^{0} + |\Psi^-\rangle^{1}
\end{equation}

where C[B+A+],\; C[B-A+],....  et al. are the CNOT operations of  A + conditioned by B+ and B- , respectively.

As shown, multiplications and addition of Eqs. 16 and 17 will result in Bell State equations which can be summerized as ,

\begin{equation}
\left.
\begin{aligned}
&|\Phi^+\rangle = |\Phi^+\rangle^{0} + |\Phi^+\rangle^{1}= B + C[B+A+] + B^{\overset{+}{+}} C[B^{\overset{+}{+}}A+] \\
&|\Phi^-\rangle = |\Phi^-\rangle^{0} + |\Phi^-\rangle^{1}= B - C[B-A+] + B^{\overset{-}{-}} C[B^{\overset{-}{-}}A+] \\
&|\Psi^+\rangle = |\Psi^+\rangle^{0} + |\Psi^+\rangle^{1}= B + C[B+A-] + B^{\overset{+}{+}} C[B^{\overset{+}{+}}A-] \\
&|\Psi^-\rangle = |\Psi^-\rangle^{0} + |\Psi^-\rangle^{1}= B + C[B-A-] + B^{\overset{-}{-}} C[B^{\overset{-}{-}}A-]
\end{aligned}
\right\} \quad |\phi\rangle
\end{equation}

where \( B^+ = |0\rangle \), \( B^{\overset{+}{+}} = |0\rangle \), \( B^- = |0\rangle \), and \( B^{\overset{-}{-}} = -|1\rangle \).
Note also that \( A^+ = |0\rangle \) and \( A^- = |1\rangle \). Note also that how \(\Phi^+\rangle^{0}\) and \(\Phi^+\rangle^{1}\) etc are formed(see Eq.22). Numerical results of Eq(22) will appear as,

\begin{equation}
\begin{aligned}
\left.
\begin{aligned}
|\Phi^+\rangle &= |0\rangle[|0\rangle] + |1\rangle[|1\rangle] \\
|\Phi^-\rangle &= |0\rangle[|0\rangle] - |1\rangle[|1\rangle] \\
|\Psi^+\rangle &= |0\rangle[|1\rangle] + |1\rangle[|0\rangle] \\
|\Psi^-\rangle &= |0\rangle[|1\rangle] - |1\rangle[|0\rangle]
\end{aligned}
\right\}
\quad
|\phi\rangle = |\Phi^+\rangle + |\Phi^-\rangle + |\Psi^+\rangle + |\Psi^-\rangle
\end{aligned}
\end{equation}

Actual implementation of the above computations using MRI pulse sequences are shown in  Fig. 15, where control pulses as a result of \( |\phi_{1}>\) and Hadamard operations and target pulses via Identity operations of \( |\phi_{2}>\) are shown.

This computation is a typical  MRI based 2-qubit Bell state formation via Hadamard and CNOT operations with which one can obtain the 4-Bell states as shown
;\( |\Phi^+\rangle,|\Phi^-\rangle,|\Psi^+\rangle,|\Psi^-\rangle\).

\subsection{An Example of the Basic Quantum Computation,\( |\Phi^+\rangle\), one of the Bell states (EPR), with MRI based Pulse Sequences}

In Fig. 16, as an example, a set of pulses that belong to the formation of the first Bell states, \( |\Phi^+\rangle\), is shown. To illustrates how these pulses control the spins and interact with target pulses in Bloch sphere, we have given two examples in Figs. 17(a) and (b) . These two figures illustrate the interactions between the control and target pulses and resulting spins in the Bloch spheres, the pulse – spin interactions of the B+C(B+A+) and \(B{\overset{+}{+}}C(B{\overset{+}{+}}A+)\)  pairs.   In Fig. 17(c), the sum of the above two are shown as a summary.  These pulses in interactions with spins in the Bloch spheres are the central part of the MRI based quantum computation.  Note the computation, the multiplication of the B+ with C(B+A+) and \(B{\overset{+}{+}}\) with \(C(B{\overset{+}{+}}A+)\)which is given by,

\begin{equation}
\begin{aligned}
|\Phi^+\rangle &= |\Phi^+\rangle^{0} + |\Phi^+\rangle^{1} = B^+ [B + A^+] + B^{\overset{+}{+}} C [B^{\overset{+}{+}} A^+] \\
&= \frac{1}{\sqrt{2}} \left( |00\rangle + |11\rangle \right)
\end{aligned}
\end{equation}

\subsection{MRI based Quantum Computing Platforms and Related Logics}
Lastly, in Figs. 18(a) and (b),  the MRI based Quantum Computing platforms and related Logic circuitry are shown for references. These illustrations show how MRI based quantum circuitry are coupled to the quantum computing system for the final computational tasks.

In Fig. 18(a), a quantum computing platform with one qubit and its couplings is shown. Subsequently, a multi-qubit platform is shown in Fig. 18(b). In Fig. 19, a brief summary diagram of the input-output couplings of the Quantum Computing platform and following Computational Quantum logics such as Pulse Sequence Generators and Computer interfaces are shown.

\section{Discussions and Summary}

In this paper, we have outlined the generation of quantum bits (qubits) using magnetic resonance imaging (MRI) techniques, distinct from previously proposed NMR-based \textsuperscript{1}H-NMR chemical shift or spectroscopic methods. We have also briefly reviewed the challenges associated with MRI-based qubit generation and related quantum computations, particularly the issue of spin purification and the implementation of MRI-compatible signal processing techniques. Notably, the bulk nature of MRI-based qubit generation presents challenges due to the ensemble or bulk spin properties, which conflict with the quantum properties required for quantum computation. Although spin purification in MRI demands rigorous and precise timing control at the picosecond (\( 10^{-12} \) seconds) level for reproducible, error-free operations, these challenges are not insurmountable, however advances in modern electronics, such as 1.5--3\,nm semiconductor chips now available in the semiconductor industry, facilitate such high-precision control.

A significant advantage of MRI-based qubit generation is its freedom from the extreme cooling requirements, such as the millikelvin (e.g., 15\,mK) temperatures needed for superconducting qubits, or the precise atomic alignment and high-vacuum conditions required for trapped-ion qubits. Given the need for multiple qubits in quantum computers to address error correction, the MRI Quantum Bit (MRQ) offers a compelling solution due to its suitability for multi-qubit generation. MRQ generation is relatively straightforward, and the resulting qubits can be operated in readily available environments. Specifically, MRI-based qubits function at room temperature, except for the main magnet (\( B_0 \)), and leverage electronics and logic developed for the MRI industry over the past four to five decades~\cite{R25,R26,R27,R28}. This wealth of accumulated experience and expertise is an invaluable resource for the future development of MRI-based quantum computers.

Recent advancements in MRI and NMR, such as the development of ultra-high-field (UHF) magnets (e.g., 28.0\,T) and high-temperature superconducting (HTS) magnets that eliminate the need for helium cooling, further enhance the prospects for MRI-based quantum computing. The increasing availability of UHF and HTS magnets further supports the pursuit and development of the MRI-based quantum computing devices proposed in this work~\cite{R30,R31}.

\section{Acknowledgements}

We would like to thank for the initial research support from NIA (Korea National Information Society Agency, NIA V-RBI-D-24047)               )  with which we have performed a number of basic qubit generation scheme using 9.4T NMR spectrometer at KAIST. We also like to thank various discussions and lectures delivered by external members : Dr. Yong Ho Lee ( Korea Research Institute of Standards and Science),
Dr. Yun Wook Chung ( Sungkyoon Kwan University), Dr. Jaewan Kim (KIAS), and Dr. Hae Woong Lee (KAIST), among others.

\section{Figures}

\begin{figure}[H]
  \centering
  \includegraphics[width=14cm]{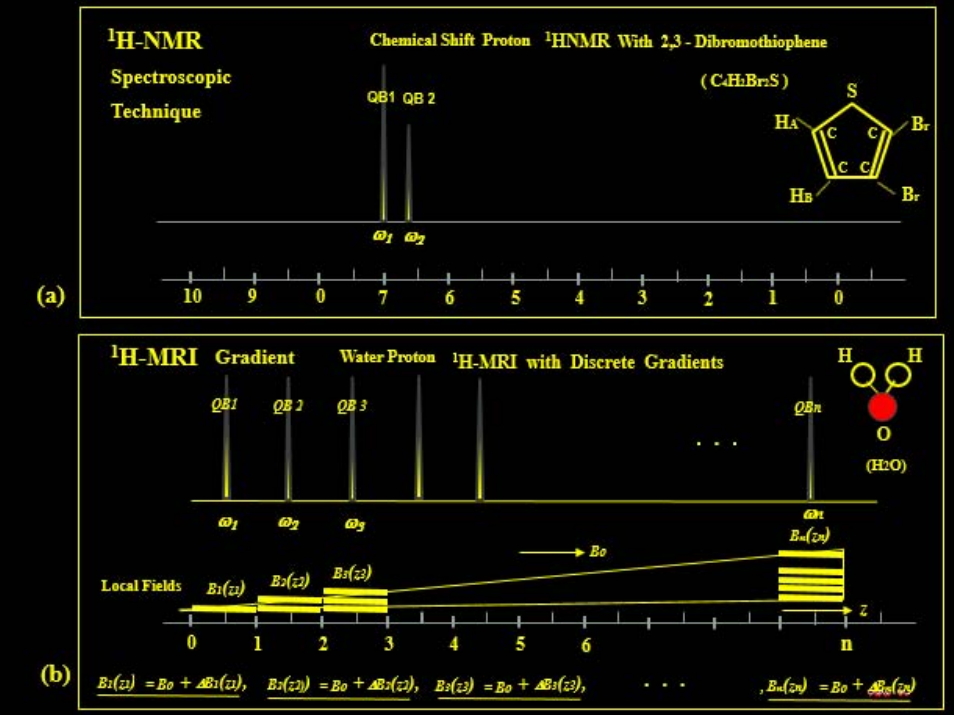}
  \caption{(a). One of the early attempts to utilize the NMR techniques for the quantum computing – the Chemical Shift  1H-NMR with 2,3-Dibromothiophene. \\
  (b). Proposed “Qubit” generation concept with  1H-MRI  using $H_2O$ and a set of MRI like Gradients (One main Gradient with a set
  of Local Reverse Gradients).}
\end{figure}

\begin{figure}[t]
  \centering
  \includegraphics[width=14cm]{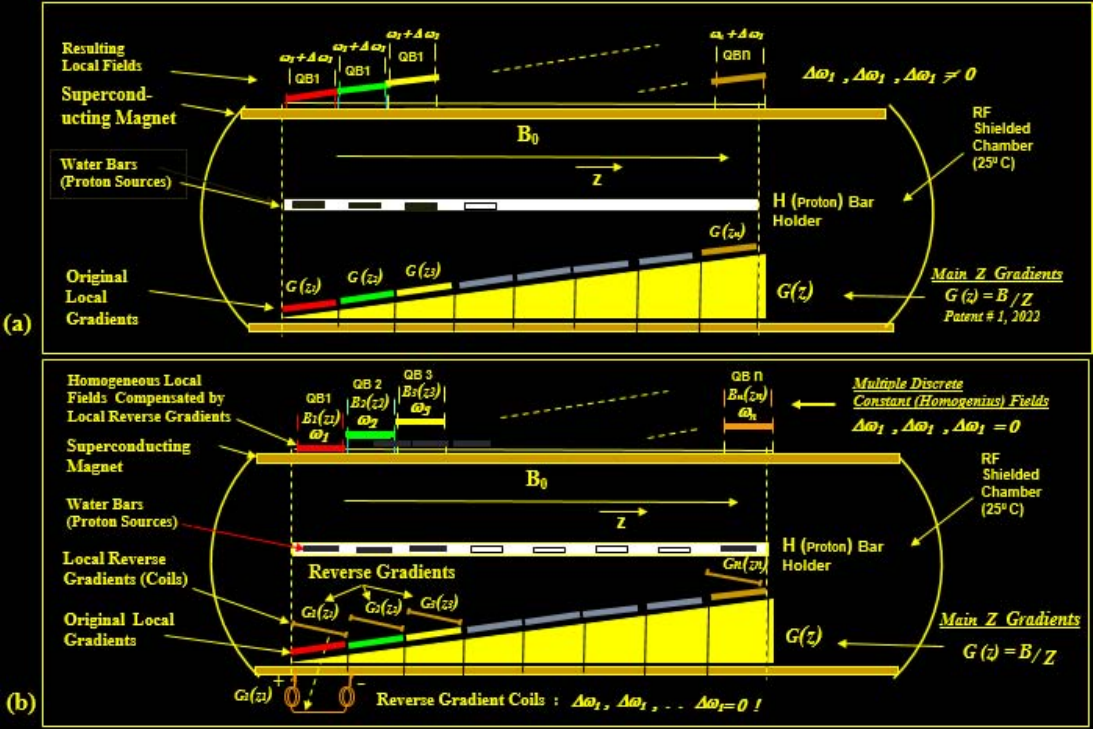}
  \caption{(a). MRI based Quantum Bit (Qubit) Generation Scheme – Original Contiguous Mode with One Main Gradient. \\
  (b). Modified MR Quantum Bit (Qubit) Generation Scheme with Multiple Local Reverse Gradients for the creation of Homogeneous Local Fields. Note the main Gradient G(z) and the other a set of small Local Reverse Gradients  $G_1(z_1)$, $G_2(z_2)$...etc  with which multiple local constant fields ( B1(z1), B2(z2), . . . )  are created.
}
\end{figure}

\begin{figure}[t]
  \centering
  \includegraphics[width=14cm]{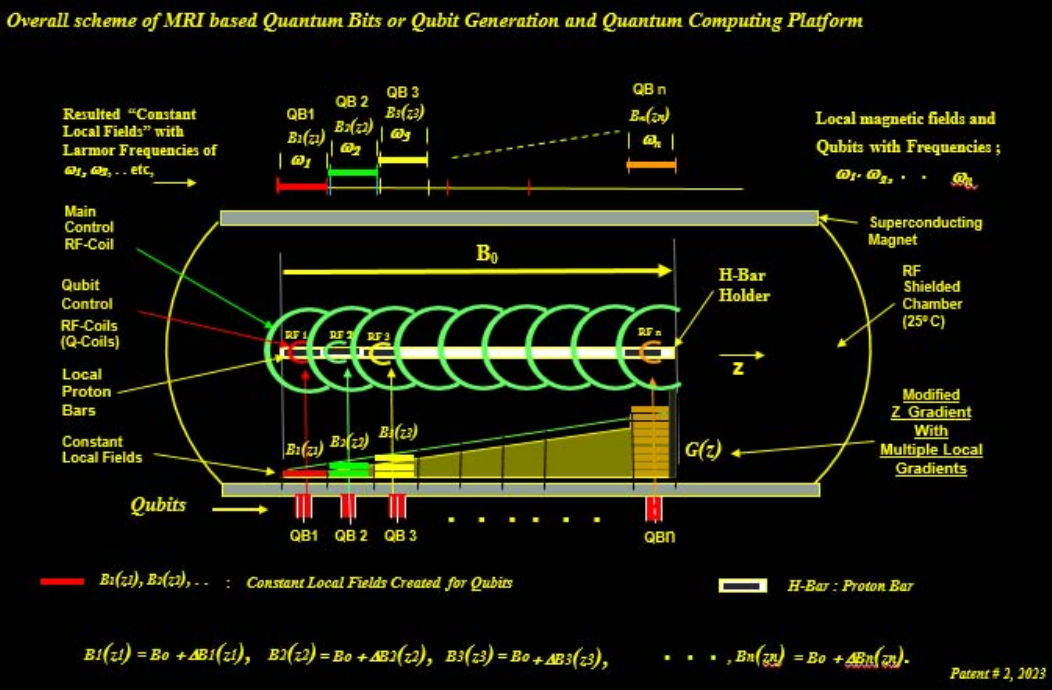}
  \caption{Overall scheme of Proposed MRI based Quantum Bits or Qubit Generation and Quantum Computing Platform. Note
  the set of small  Qubit control RF Coils ( RF1, RF2,  .  .  .) and  Local Proton Bars inside the Qubit coils. Both of them are
  located inside the Large Main Control RF Coil (green) .  All the Gradients and RF Coils are positioned  inside the main
  magnet similar way to  the  conventional MRI Scanner.}
\end{figure}

\begin{figure}[t]
  \centering
  \includegraphics[width=14cm]{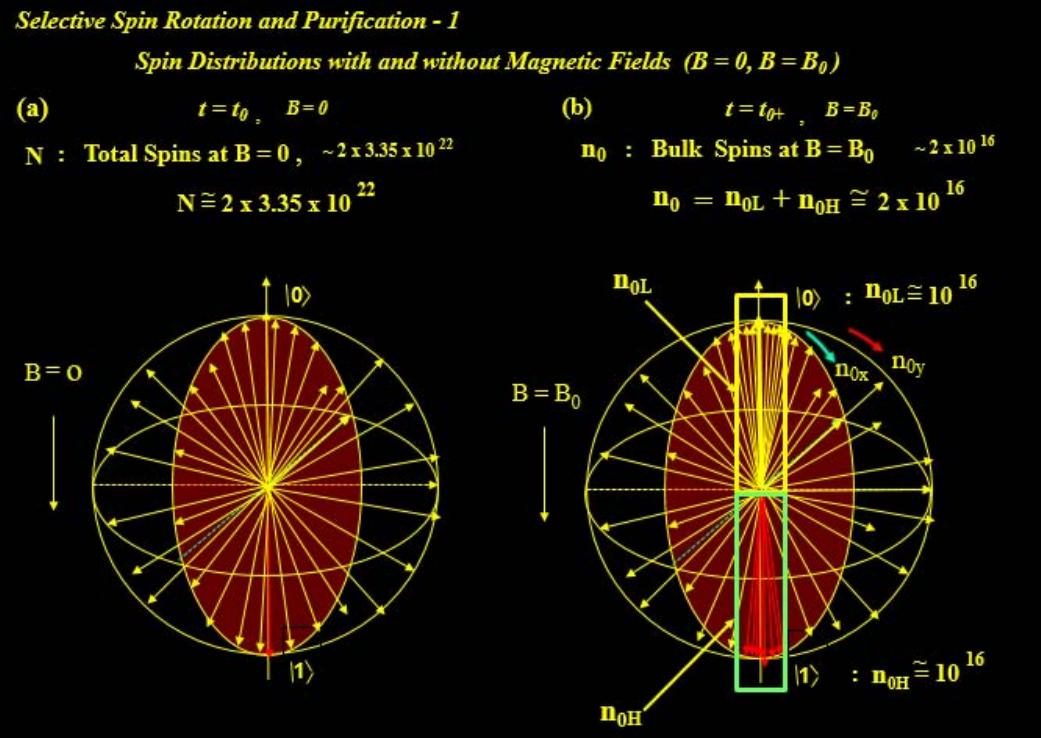}
  \caption{Basic concept illustrated for the “Spin Purification” process with MRI or NMR techniques.  (a). and (b)  are with and without magnetic field Bo , respectively. These two states we have defined as “Total Spins” and “Bulk Spins”, respectively.}
\end{figure}

\begin{figure}[t]
  \centering
  \includegraphics[width=14cm]{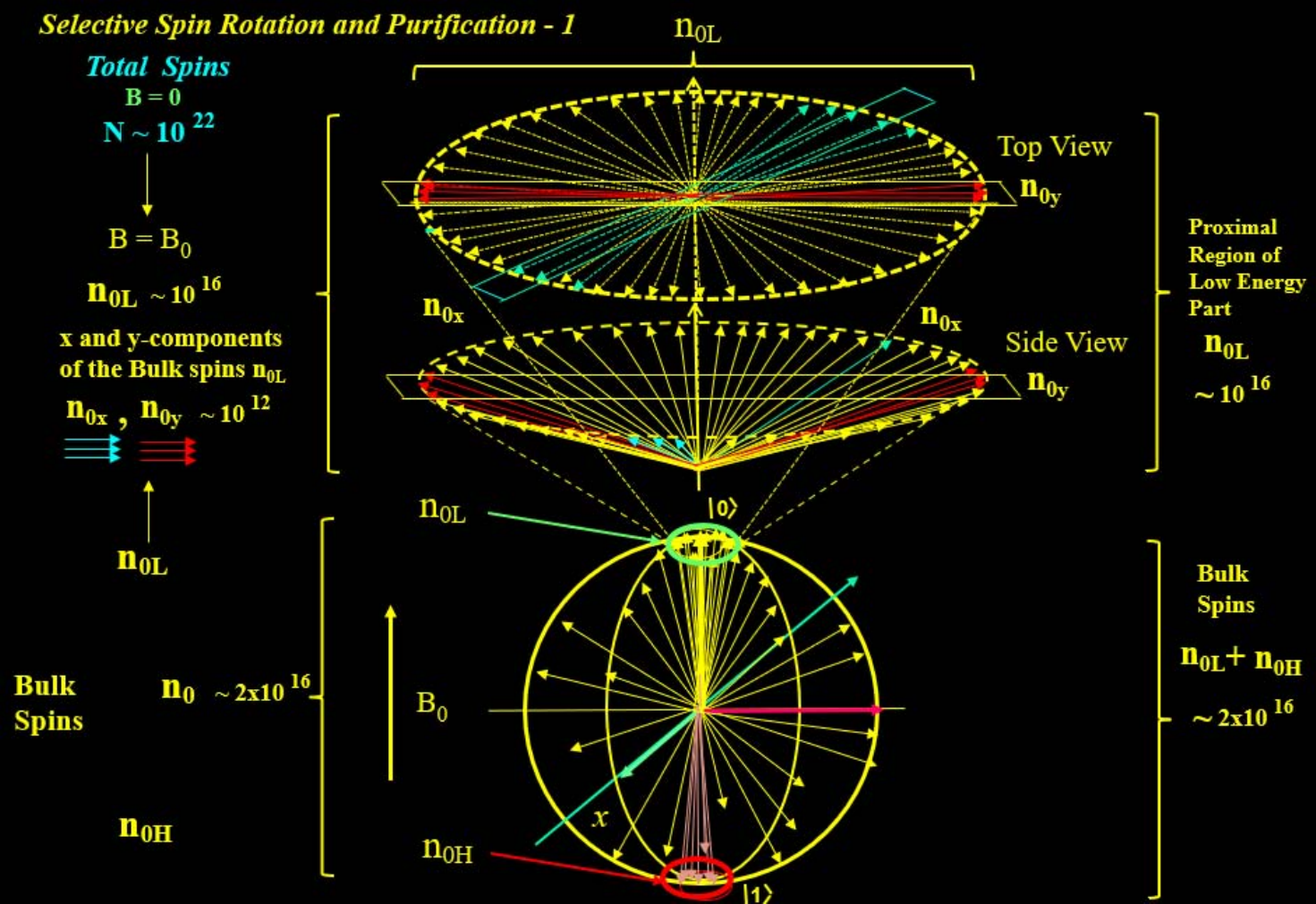}
  \caption{Spin Purification Technique  : The  Bulk Spins which can be Purified by various Purification techniques. Note \( n_{\text{Ox}} \) and \( n_{\text{Oy}} \), are the x and y-components of the Bulk spins \( n_{\text{OL}} \). These two components are the spin groups that are useful for MRI based quantum computing.}
\end{figure}

\begin{figure}[t]
  \centering
  \includegraphics[width=14cm]{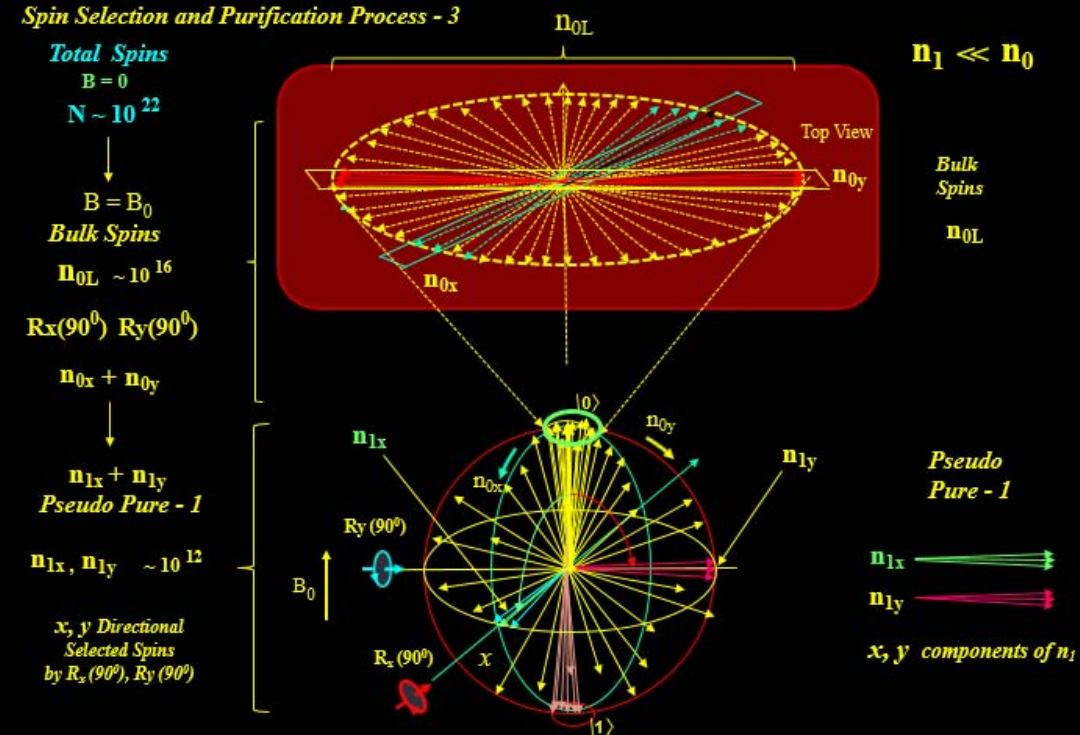}
  \caption{An example of a simple Spin Purification Technique : From Bulk to “Pseudo-Pure” states via $90^\circ$ Rotation RF Pulses.  The selected or purified spin groups, \( n_{\text{1x}} \) and \( n_{\text{1y}} \) from the Bulk spins \( n_{\text{OL}} \) are illustrated as an example. Results of these spin purifications via $90^\circ$ rotations, \( n_{\text{1x}} \) and \( n_{\text{1y}} \), we have coined as “Pseudo Pure-1” state spins.}
\end{figure}

\begin{figure}[t]
  \centering
  \includegraphics[width=14cm]{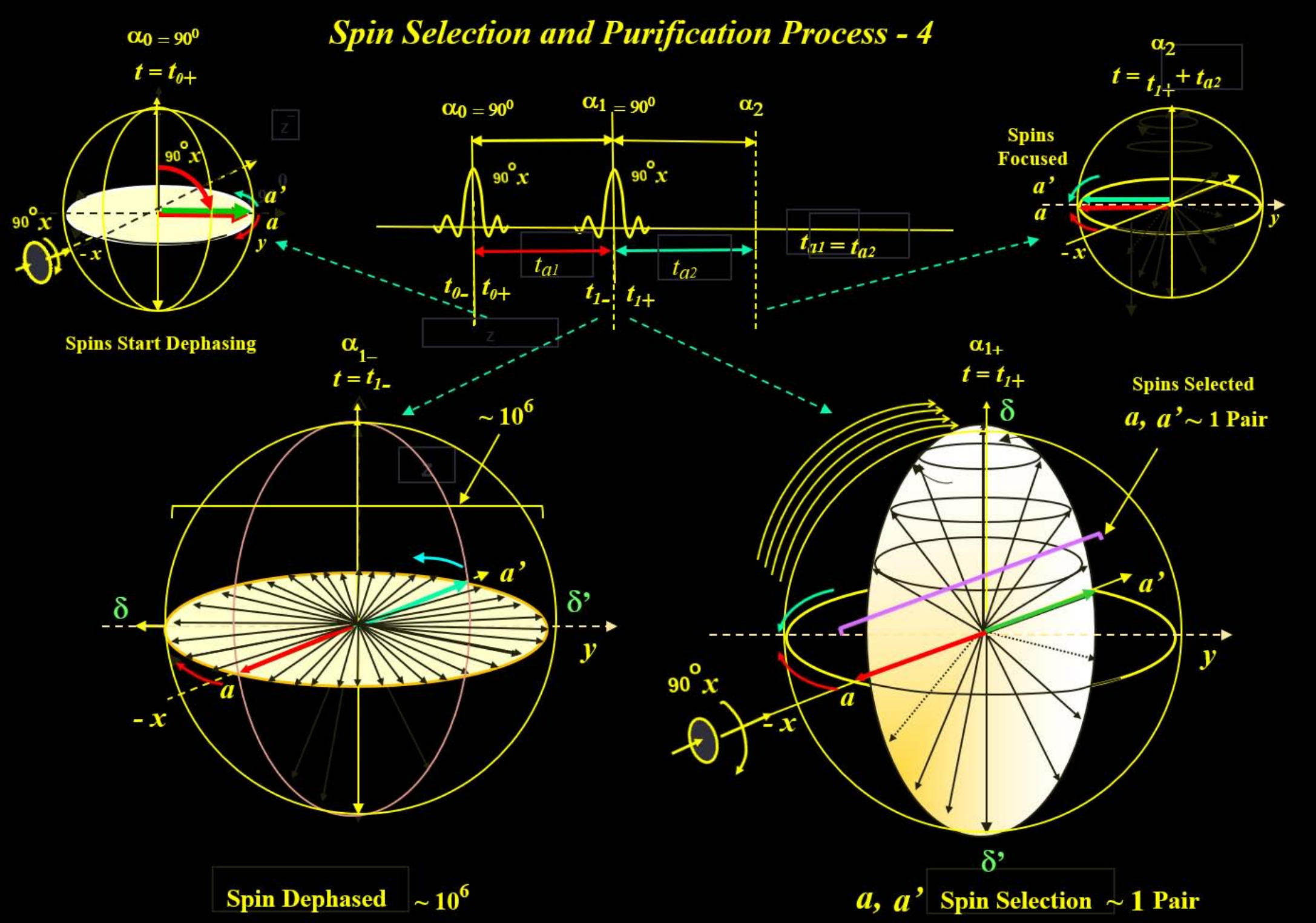}
  \caption{An example of a more advanced “Spin Purification”  technique via selection of a spin pair a-a’ at a dephased state  by a $90^\circ_x$ rotation with precision timings,  $t_{a1}$ and $t_{a2}$ .}
\end{figure}

\begin{figure}[t]
  \centering
  \includegraphics[width=14cm]{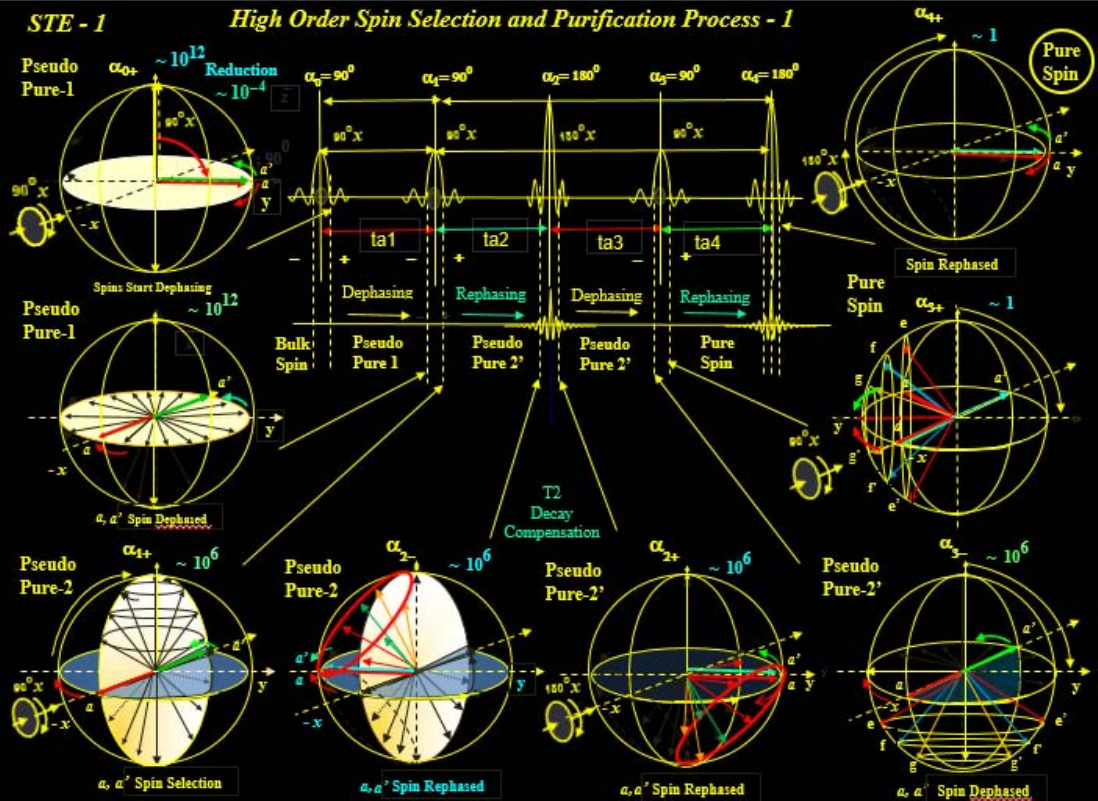}
  \caption{Another  Purification technique using multiple $90^\circ_x$ spin rotations, known as  “Modified Stimulated Echo (STE)” technique using two $180^\circ$ spin echos.
  Note the multiple $90^\circ$ rotations effectively select the a-a’ spin pairs and eventually lead to the “Pure” state. Here, we simply
  assumed each purification factor as $10^{-6}$.  To successfully achieve high degree of Purification, however, it is important to
  control the time intervals between the $90^\circ$ RF pulses, the ta1, ta2, … , as well as environmental noises which will  affect the
  spin locations at the nominal positions, for example, the spin pair (a, a’)  at x-axis. Additional two $180^\circ$ rotations; These additions will compensate the errors that might occur during the dephasing and rephasing
  periods (Ta1, Ta2, Ta3  and Ta4) of the spin pair (a, a’).}
\end{figure}

\begin{figure}[t]
  \centering

  \begin{subfigure}[t]{14cm}
    \centering
    \includegraphics[width=14cm]{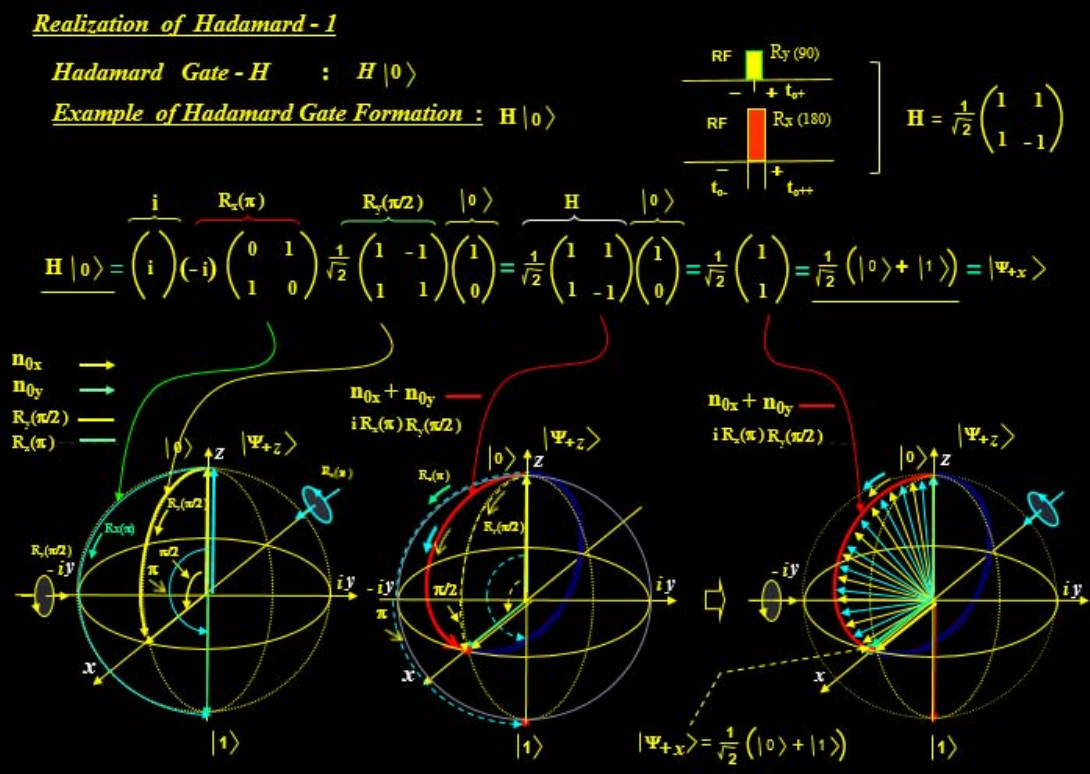}
    \caption{Spin trajectory of Hadamard operation with \(|0\rangle\). Note the summed trajectory (red line) resulted from simultaneous activations of \(n_{0x}\) and \(n_{0y}\).}
    \label{fig:fig9a}
  \end{subfigure}
  \vskip\baselineskip

  \begin{subfigure}[t]{14cm}
    \centering
    \includegraphics[width=14cm]{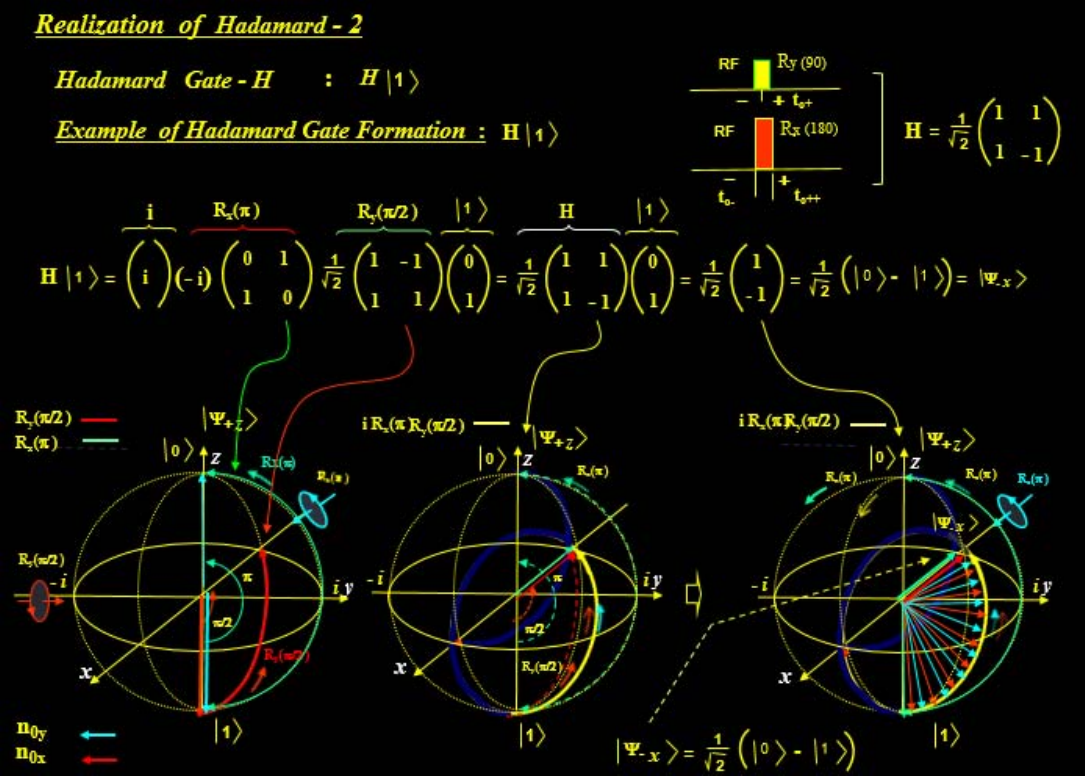}
    \caption{Same as (a), but for \(|1\rangle\).}
    \label{fig:fig9b}
  \end{subfigure}

  \caption{Spin trajectories during Hadamard gate operations.}
  \label{fig:fig9}
\end{figure}

\begin{figure}[t]
  \centering
  \includegraphics[width=14cm]{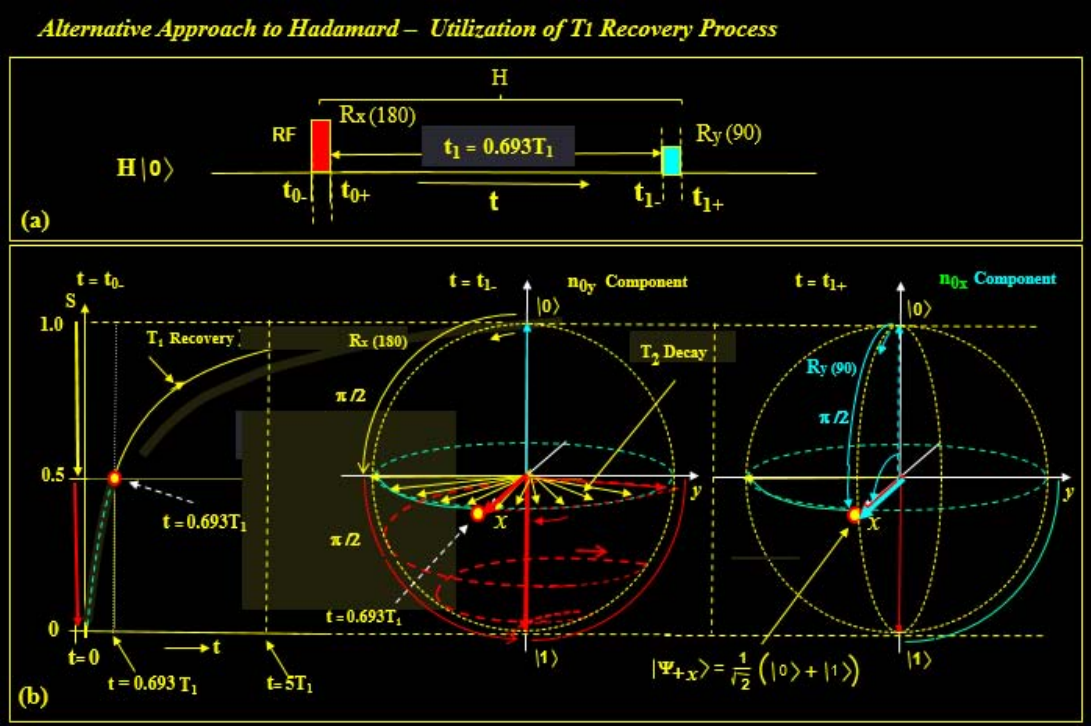}
  \caption{(a). Pulse sequence of the proposed alternative form of the  Hadamard Gate using the T1 recovery process together with a simultaneous application of $90^\circ$y pulse in coincidence with \( n_{\text{1y}} \) component. \\
  (b). T1 recovery process and spin recovery trajectory in Bloch sphere. Note the time, t1 = 0.693 T1,  which we have used for the Time-Coincidence with the second RF pulse , Ry(90), for the \( n_{\text{Ox}} \) component.}
\end{figure}

\begin{figure}[t]
  \centering
  \includegraphics[width=14cm]{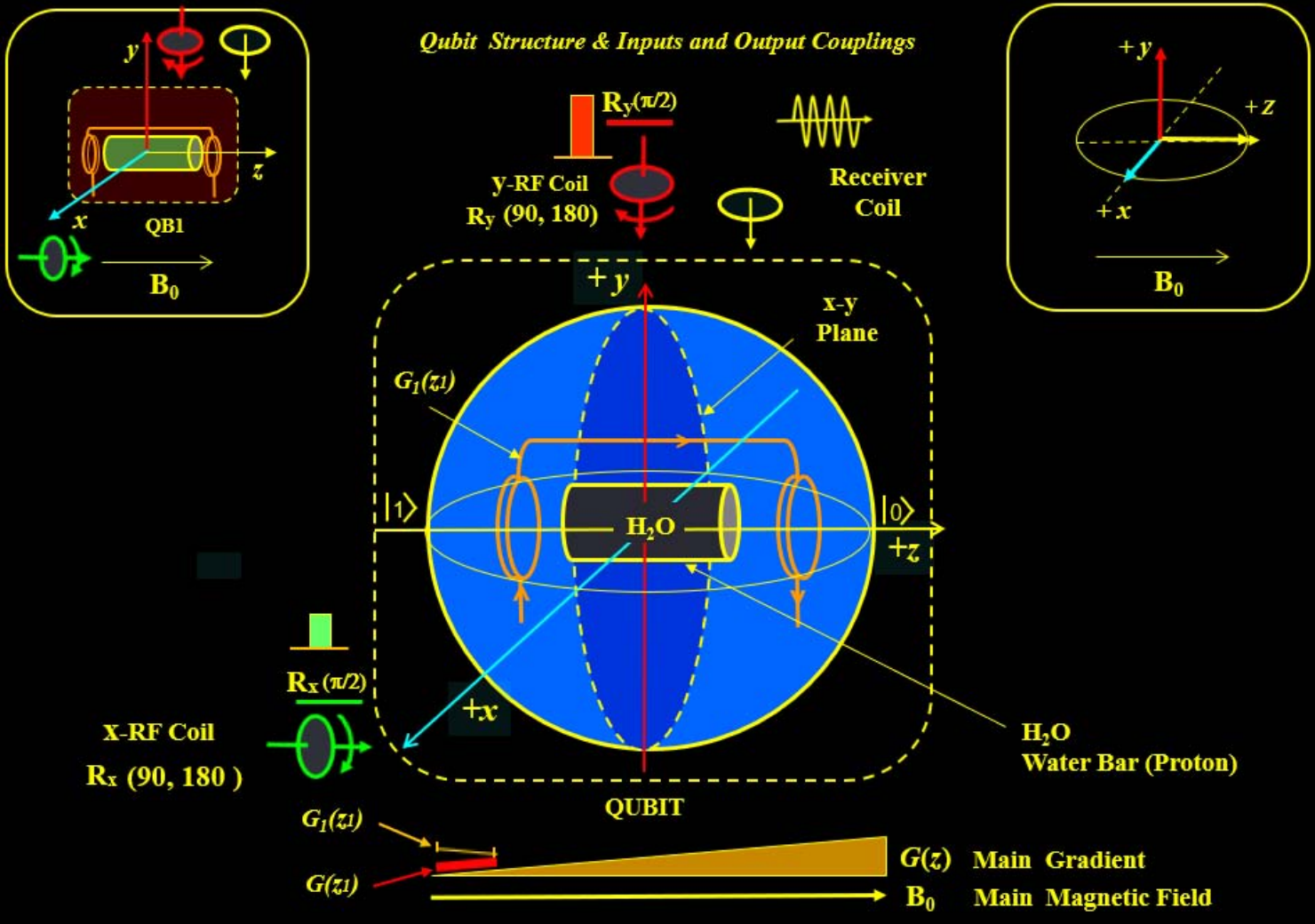}
  \caption{MR Qubit and Input-Output Coupling Diagram that can be applied to MRI based Quantum Computation. Note the two control inputs, x and y RFs and a receiver coil for output signal readout.}
\end{figure}

\begin{figure}[t]
  \centering
  \includegraphics[width=14cm]{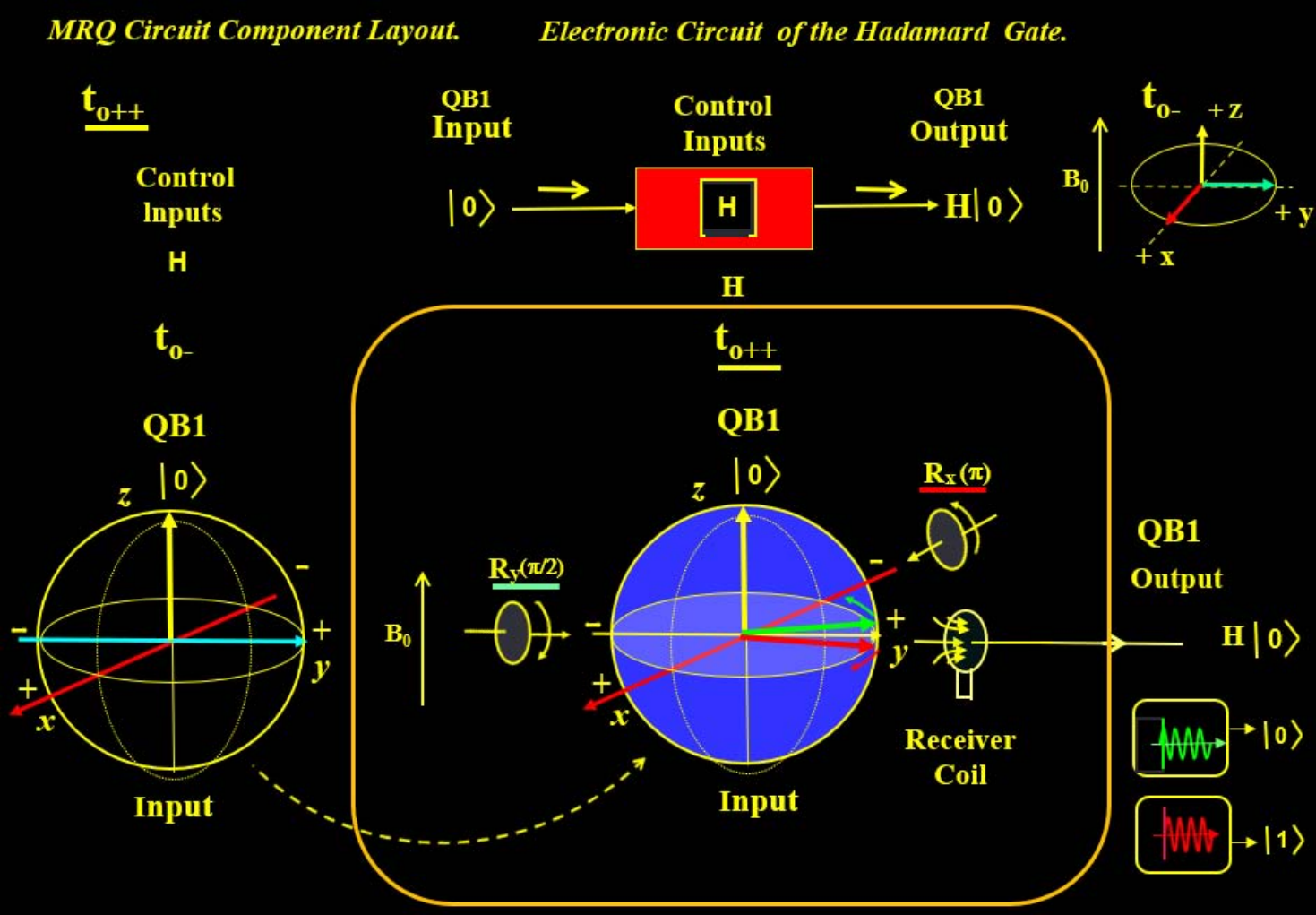}
  \caption{Qubit and input-output coupling with circuit diagram for the Hadamard gate operation at the initial state, \( t = t_0 \), and \( t = t_0^++ \). Note the directional change on $B_0$.}
\end{figure}

\begin{figure}[t]
  \centering
  \includegraphics[width=14cm]{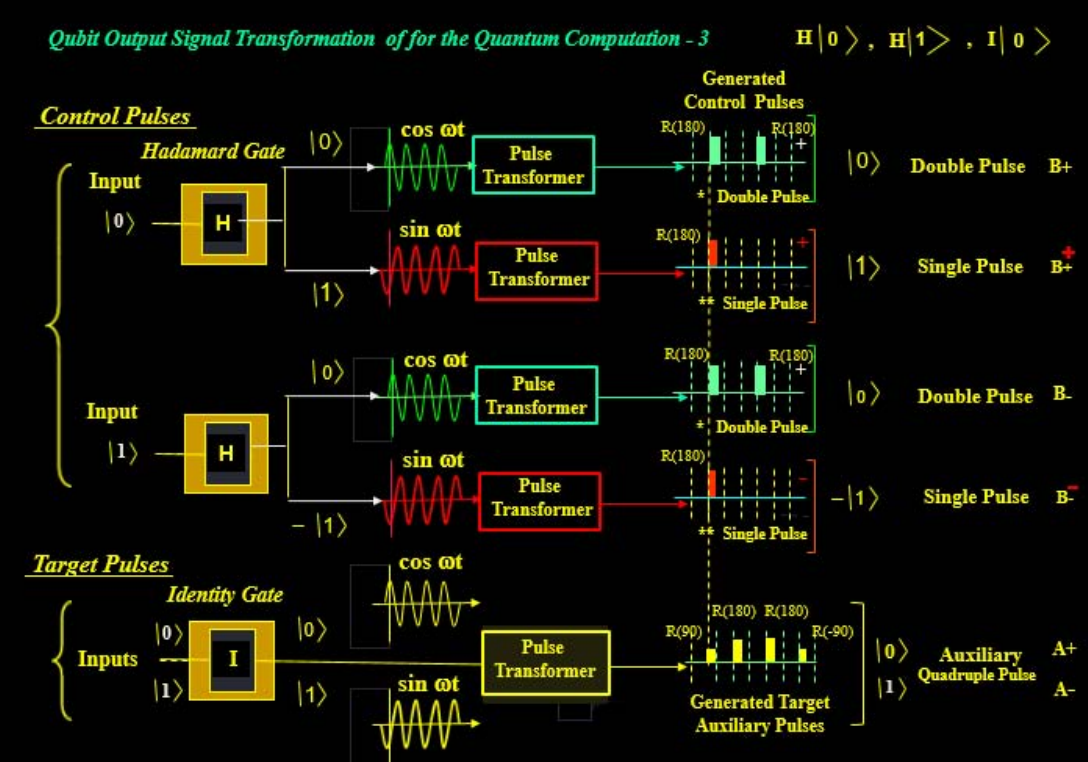}
  \caption{Block diagram of the MRI Compatible Pulse Transformation  of  the Control and Target signals for the quantum signal processing applications. Note that the R(180) and R(90) are the RF pulses of $180^\circ$ and $90^\circ$, respectively}
\end{figure}

\begin{figure}[t]
  \centering

  \begin{subfigure}[t]{14cm}
    \centering
    \includegraphics[width=14cm]{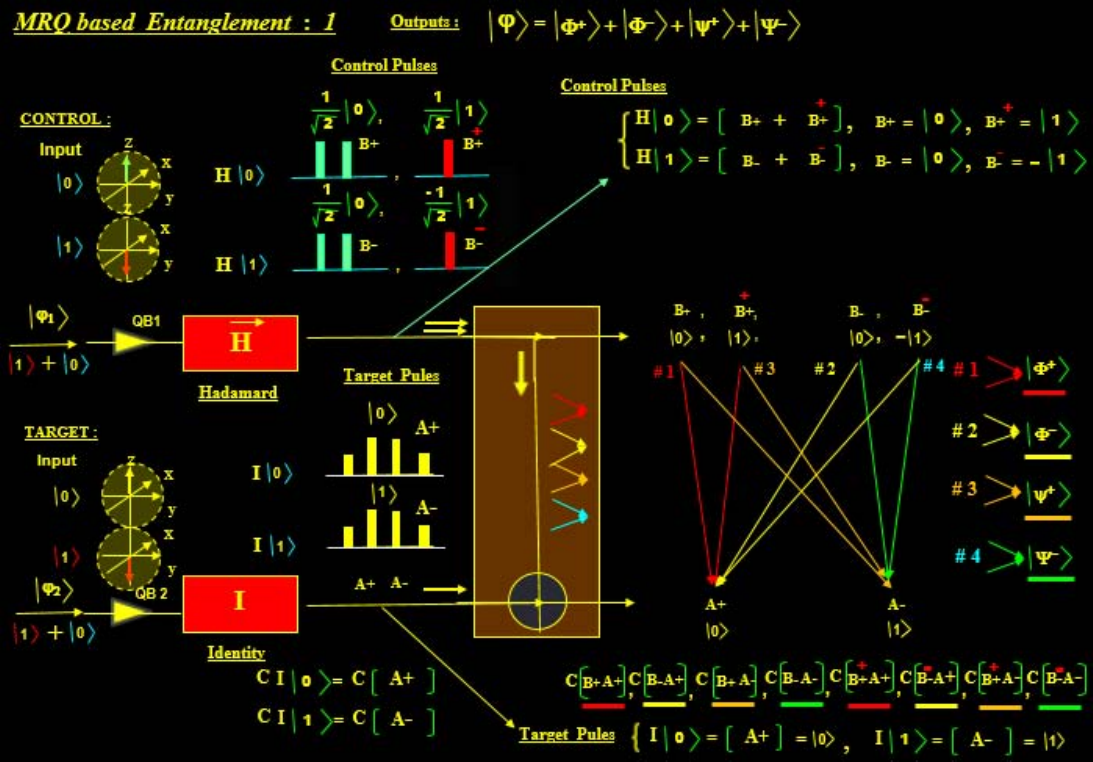}
    \caption{Block diagram of the MRI-based realization of "Entanglement" using pulses generated by the transformer.}
    \label{fig:fig14a}
  \end{subfigure}
  \vskip\baselineskip

  \begin{subfigure}[t]{14cm}
    \centering
    \includegraphics[width=14cm]{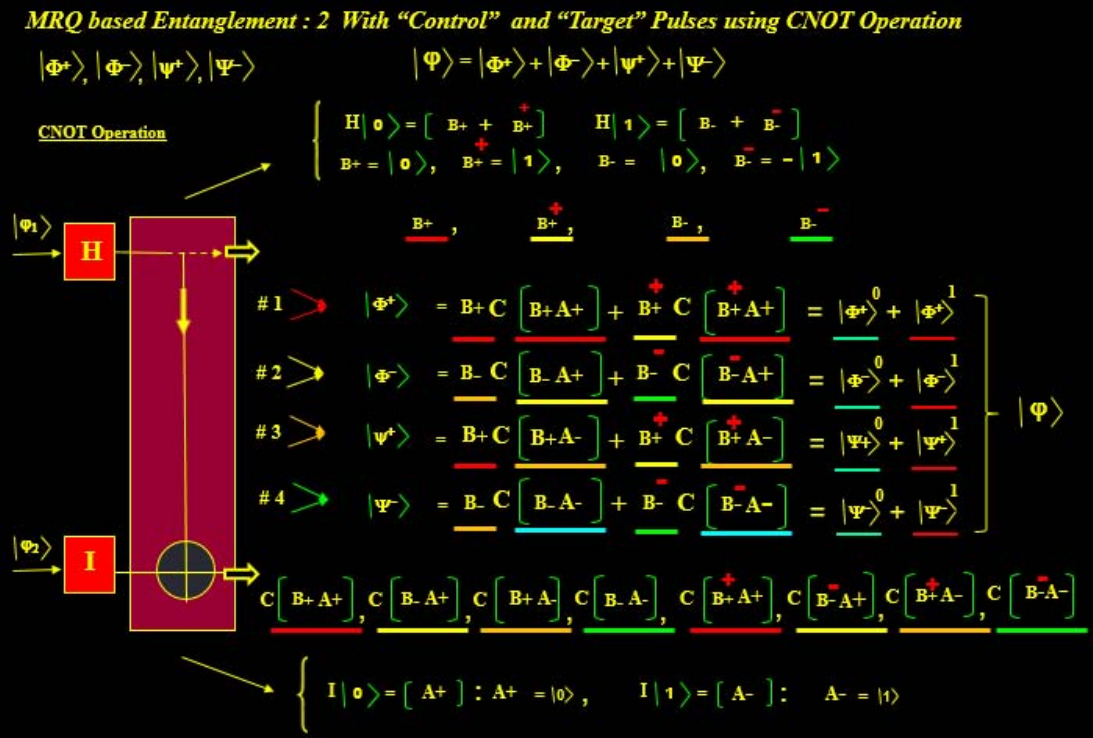}
    \caption{Numerical example of the MRI-based realization of "Entanglement" using pulses generated by the transformer.}
    \label{fig:fig14b}
  \end{subfigure}

  \caption{MRI-based realization of "Entanglement" using transformer-generated pulses.}
  \label{fig:fig14}
\end{figure}

\begin{figure}[t]
  \centering
  \includegraphics[width=14cm]{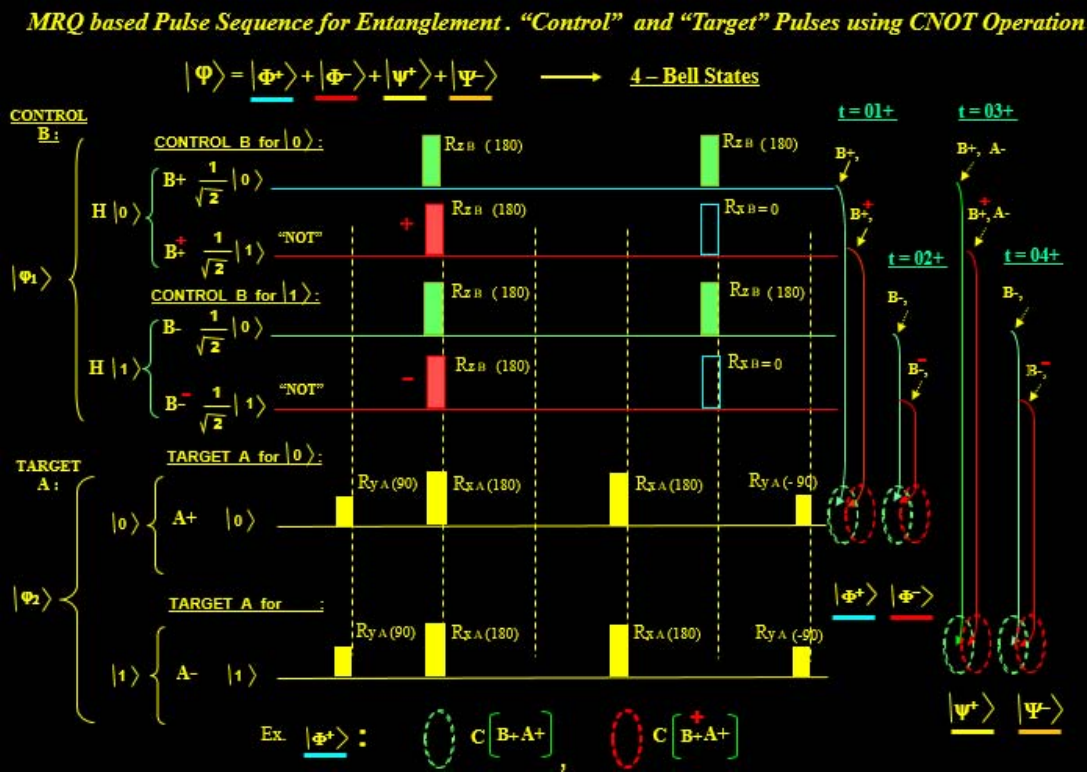}
  \caption{Pulse Sequence of the MRI based realization of "Entanglement" using the Pulses generated by the Transformer. Note the
     formation of the 4-Bell States. Note that the \( \phi_1 \)and\( \phi_2\) are the Control and Target inputs, respectively. Note also that the $R_{zB}(180)$ can be replaced by $R_{yB}(180)$.}
\end{figure}

\begin{figure}[t]
  \centering
  \includegraphics[width=14cm]{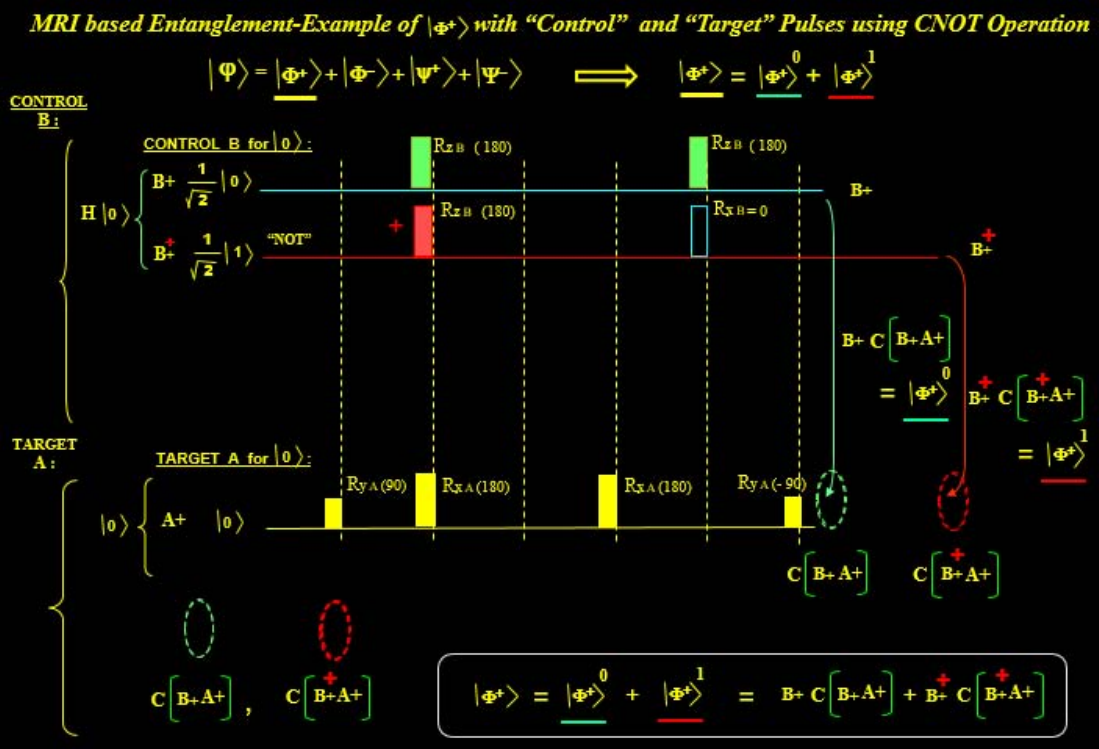}
  \caption{Example of the Pulse Sequence of the MRI based realization of "Entanglement" using the Pulses generated by the Transformer. This part of the Bell state is the first part of \( |\phi> \), the \( |\Phi^+> \).  Note that how CNOT operation is performed.}
\end{figure}

\begin{figure}[t]
  \centering
  \includegraphics[width=14cm]{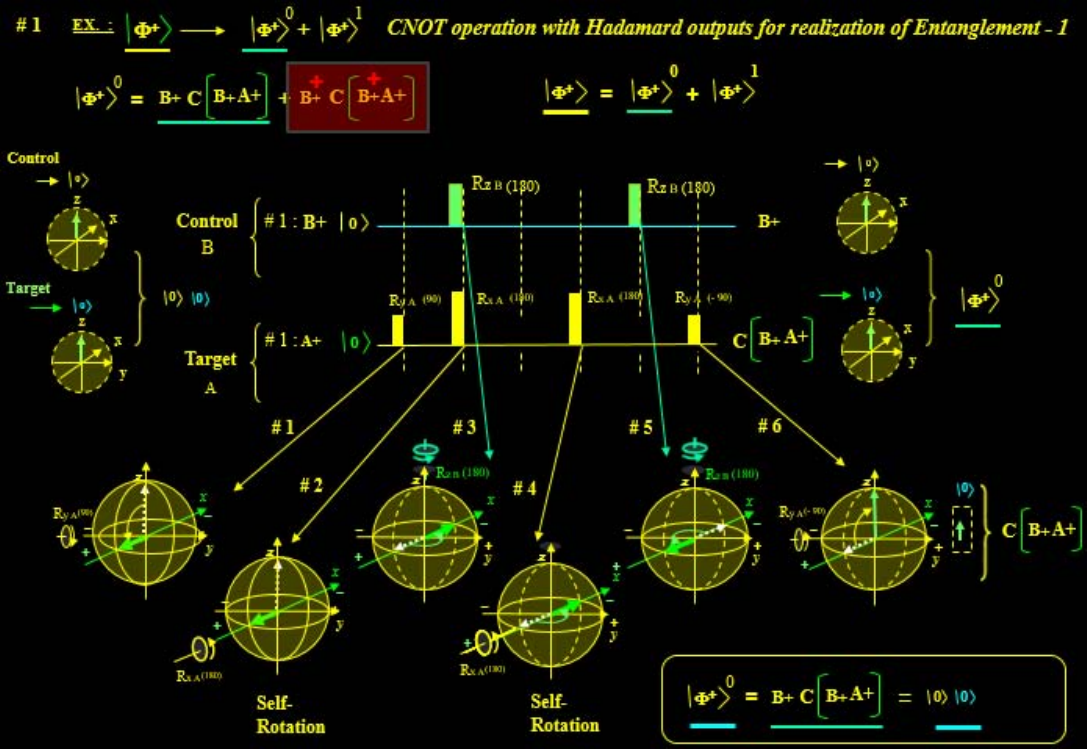}
  \caption{(a) MRI-based Pulse Sequence for the realization of \( |\Phi^+> \) and \( |\Phi^+>^0 \). Note how the CNOT operation is performed in the Bloch sphere by the Control and Target pulses.}
  \label{fig:fig17a}
\end{figure}

\begin{figure}[t]
  \ContinuedFloat
  \centering
  \includegraphics[width=14cm]{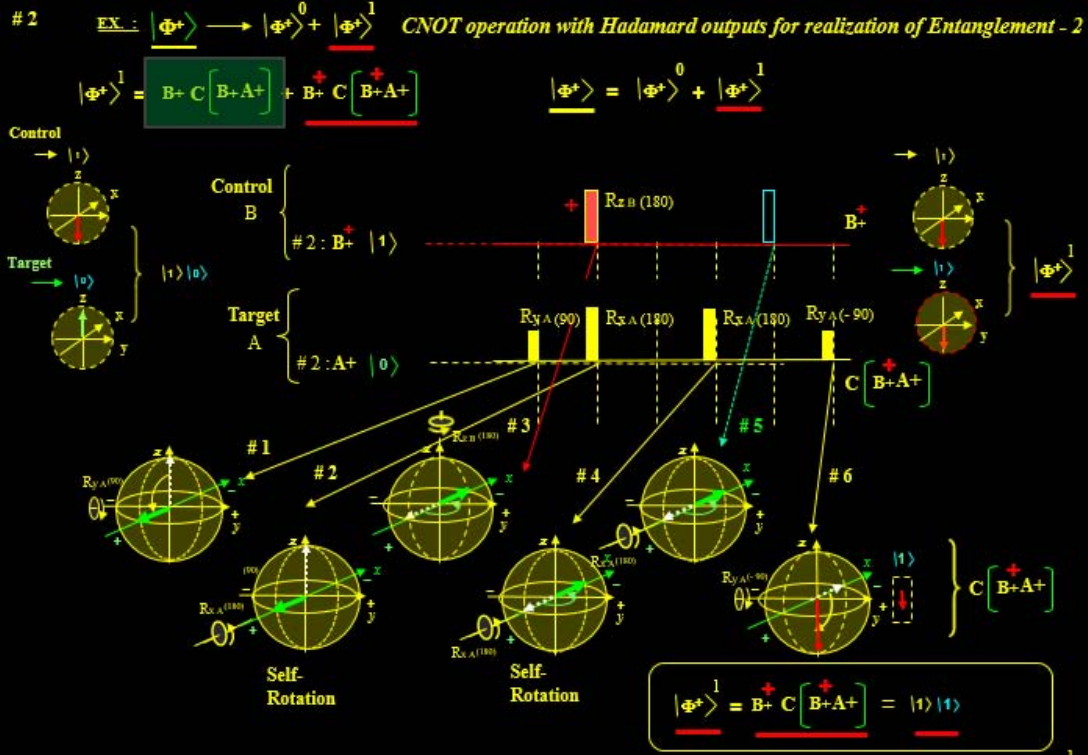}
  \caption{(b) MRI-based Pulse Sequence for the realization of \( |\Psi^+>^1 \).Note how the CNOT operation is performed in the Bloch sphere by the Control and Target pulses.}
  \label{fig:fig17b}
\end{figure}

\begin{figure}[t]
  \ContinuedFloat
  \centering
  \includegraphics[width=14cm]{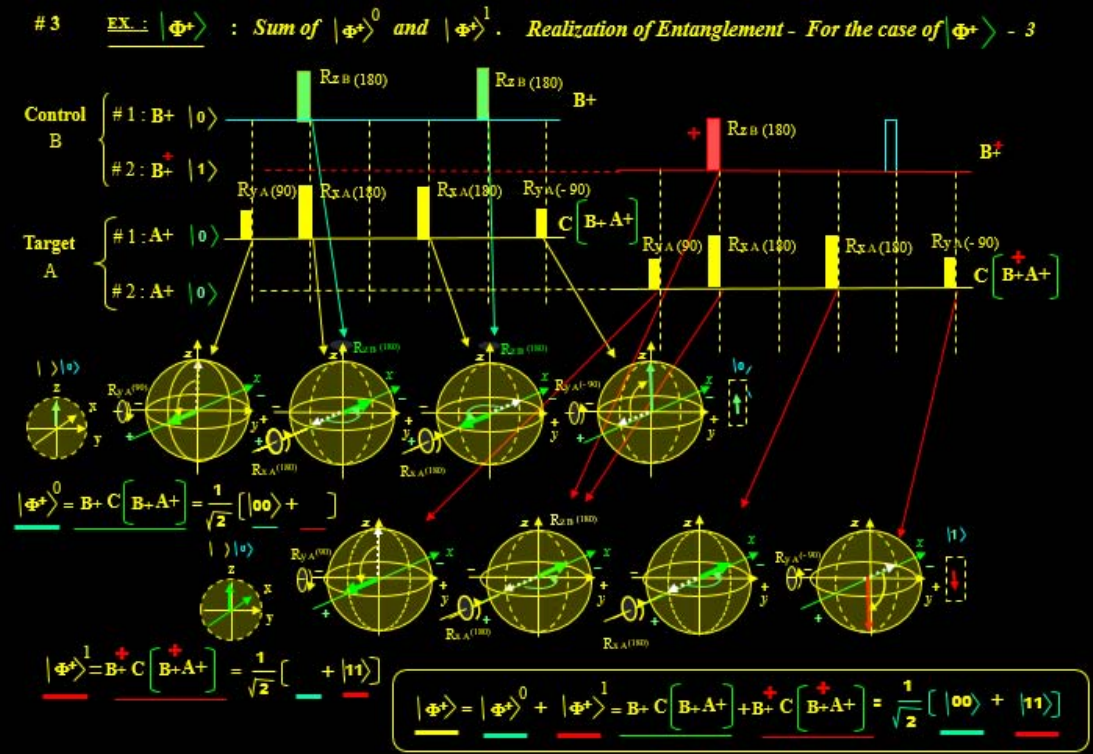}
  \caption{(c) Sum of the (a) and (b), the Bell states \( |\Phi^+> = |\Phi^+>^0 +|\Phi^+>^1 \).}
  \label{fig:fig17c}
\end{figure}

\begin{figure}[t]
  \centering

  \begin{subfigure}[t]{14cm}
    \centering
    \includegraphics[width=14cm]{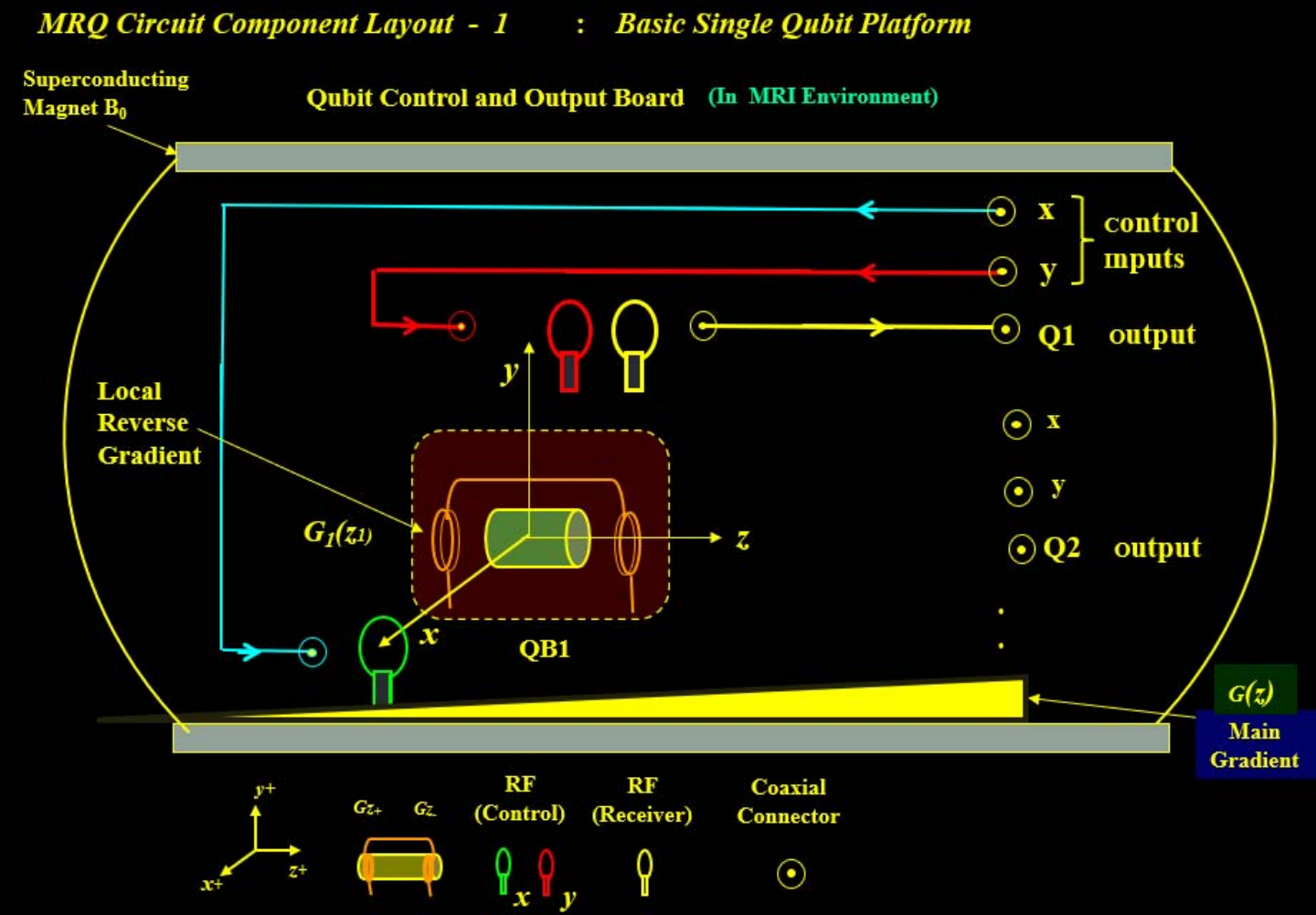}
    \caption{One Qubit Quantum Computing Platform with two Control Inputs and one Output Coupling.}
    \label{fig:fig18a}
  \end{subfigure}
  \vskip\baselineskip

  \begin{subfigure}[t]{14cm}
    \centering
    \includegraphics[width=14cm]{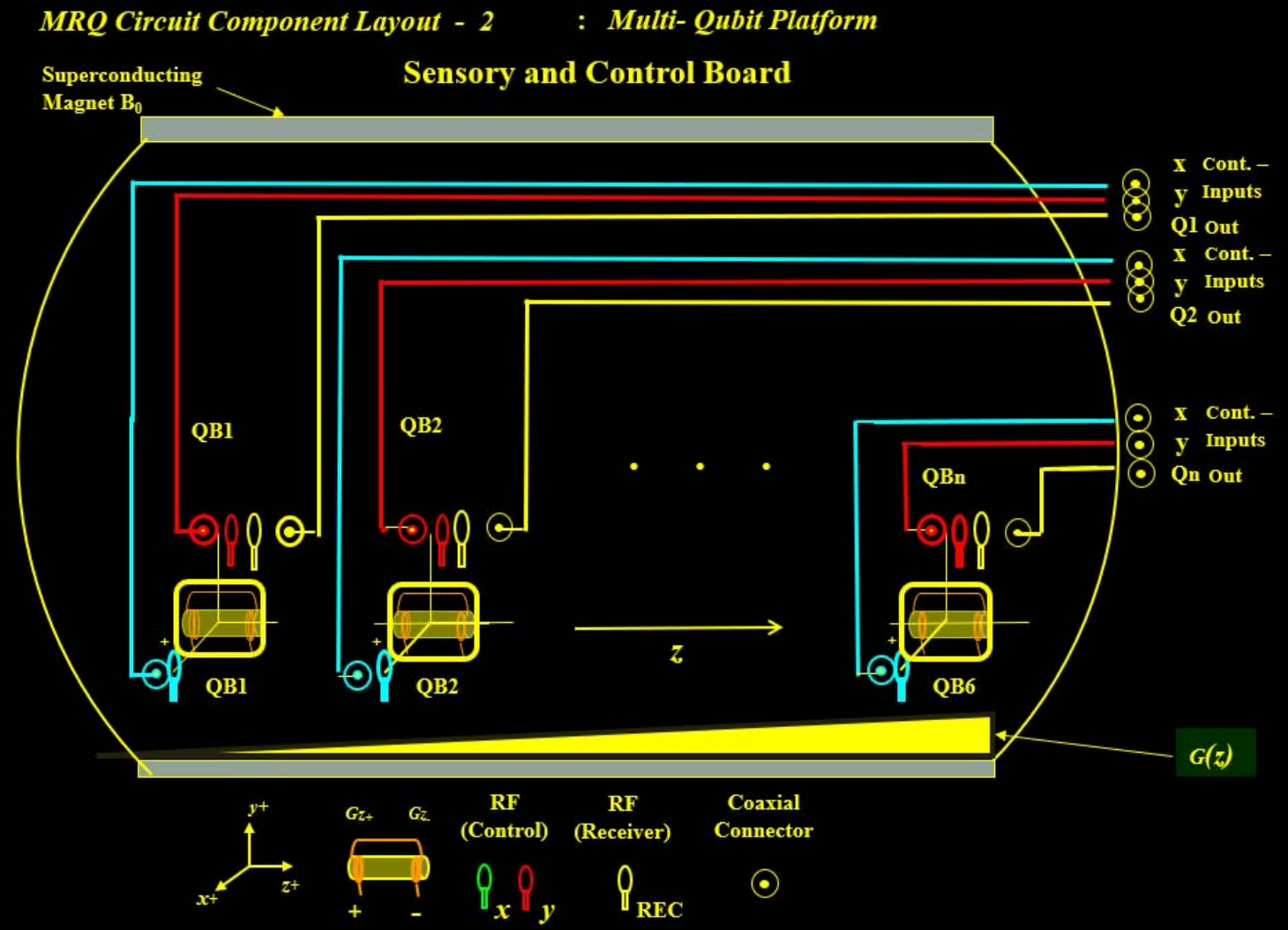}
    \caption{Quantum Computing Platform with multiple qubits and Input and Output Couplings.}
    \label{fig:fig18b}
  \end{subfigure}

  \caption{Illustration of quantum computing platforms with different qubit configurations.}
  \label{fig:fig18}
\end{figure}

\begin{figure}[t]
  \centering
  \includegraphics[width=14cm]{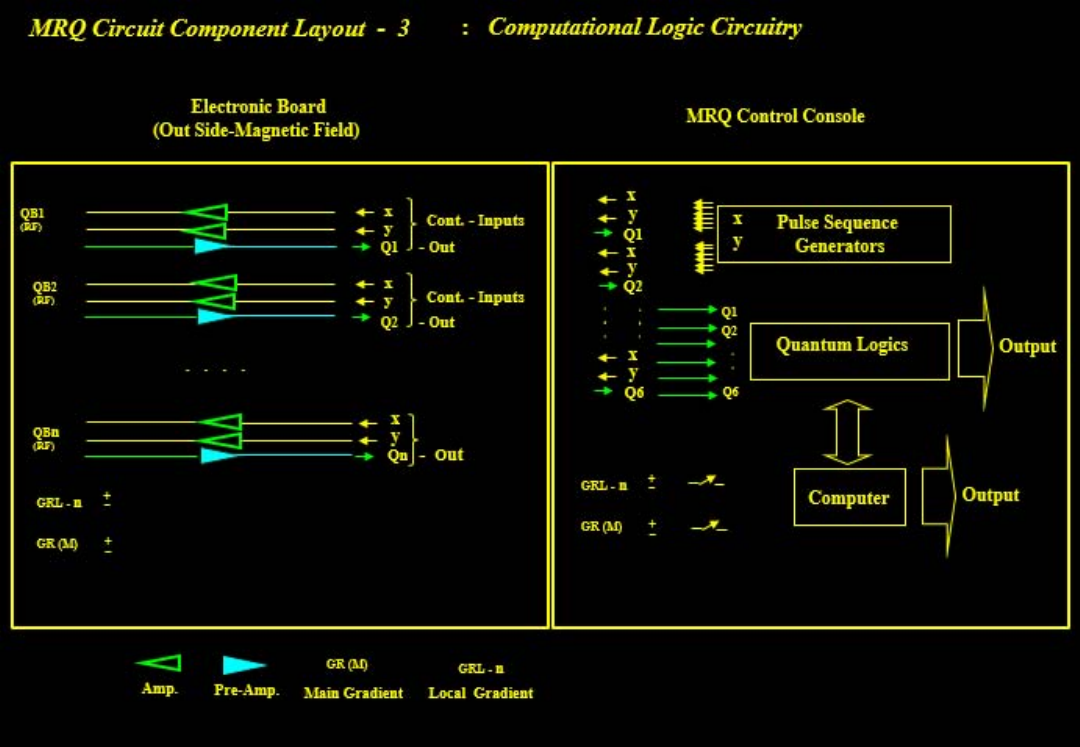}
  \caption{Quantum Computing System with Logic Circuits and Accessary Units.}
\end{figure}

\clearpage
\appendix
\begin{appendices}
\counterwithin{figure}{section}
\section{Appendix}

\begin{figure}[H]
  \centering
  \includegraphics[width=14cm]{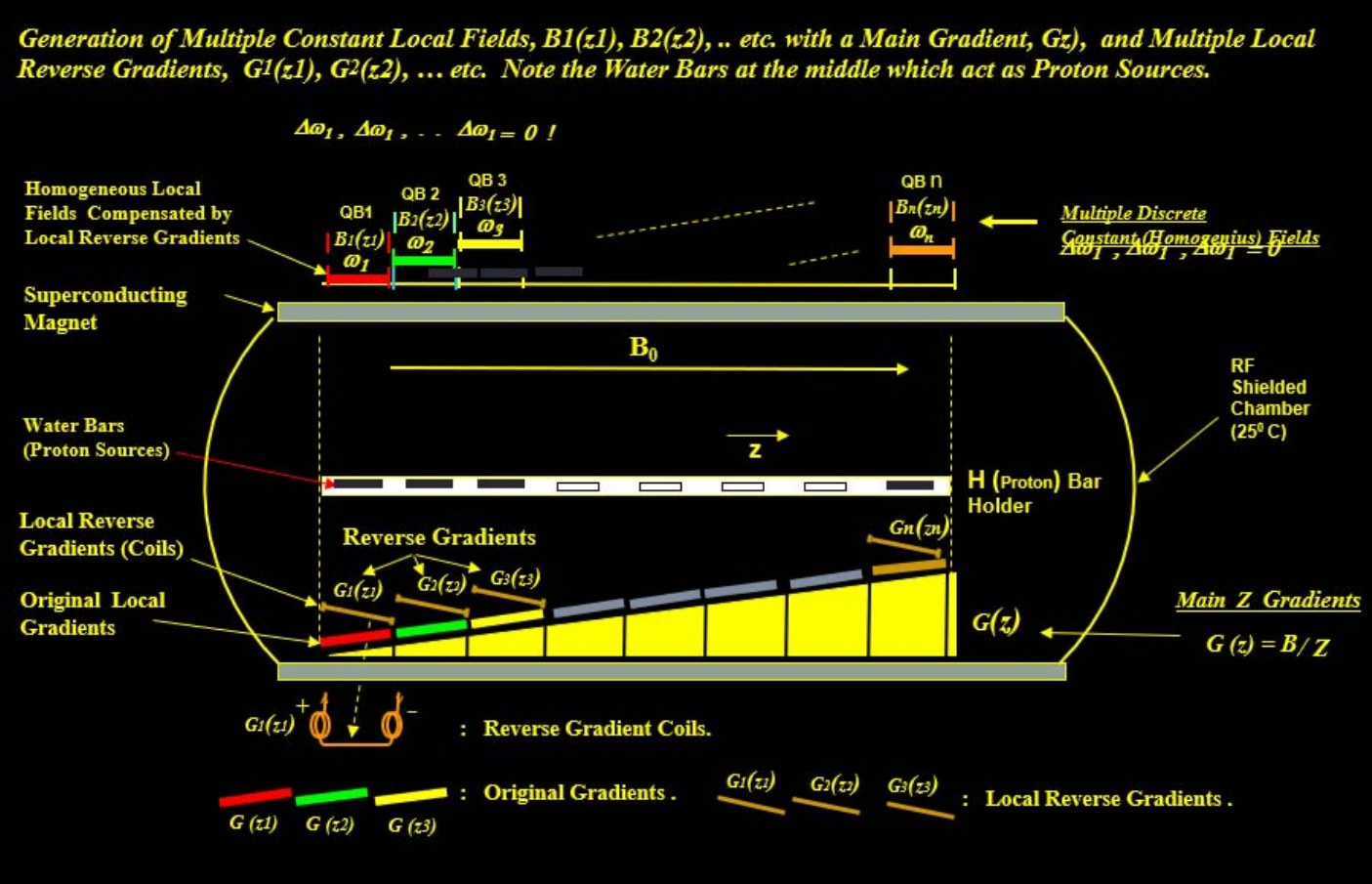}
  \caption{MR Quantum Bit (Qubit) Generation Scheme with Multiple Local Reverse Gradients for the creation of Multiple Homogeneous Local Fields. Note the main Gradient $G(z)$ and the other set of small Local Reverse Gradients $G_1(z_1), \ldots$ with which multiple local constant fields ($B_1(z_1), B_2(z_2), \ldots$) are created. Note the $\Delta\omega_1$, $\Delta\omega_2$, \ldots, $\Delta\omega_n = 0$. MRQ: MRI-based Quantum Bit.}
\end{figure}

\begin{figure}[t]
  \centering
  \includegraphics[width=14cm]{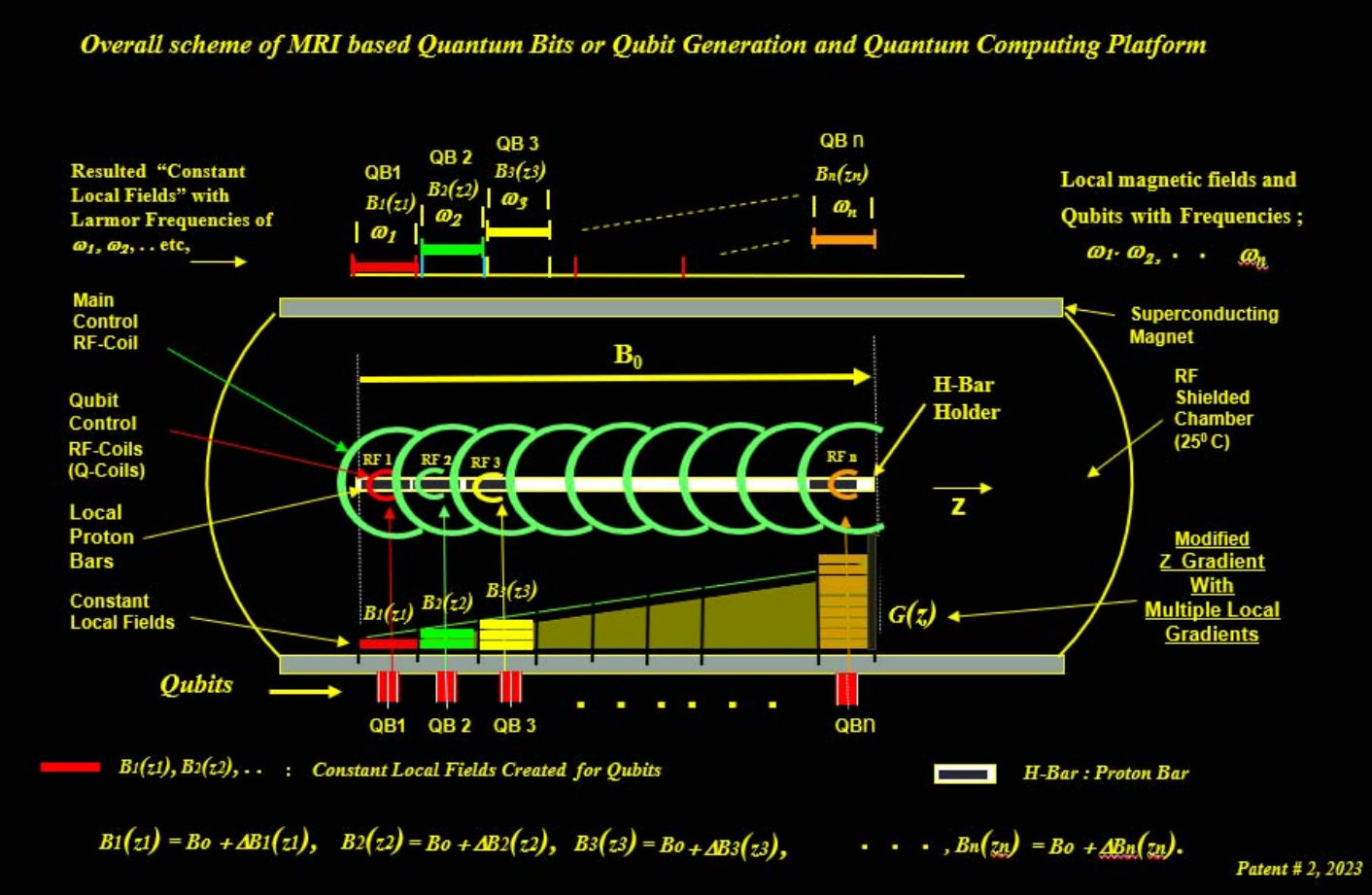}
  \caption{Overall scheme of MRI-based Quantum Bits or MRQ Generation and Quantum Computing Platform. Note the set of small Qubit control RF Coils (Q-Coils, RF1, RF2, \ldots) and Local Proton Bars inside the Q-Coils. Both are located inside the Large Main Control RF Coil (green). All the Gradients and RF Coils are positioned inside the main magnet, similar to a conventional MRI Scanner.}
\end{figure}

\begin{figure}[t]
  \centering
  \includegraphics[width=14cm]{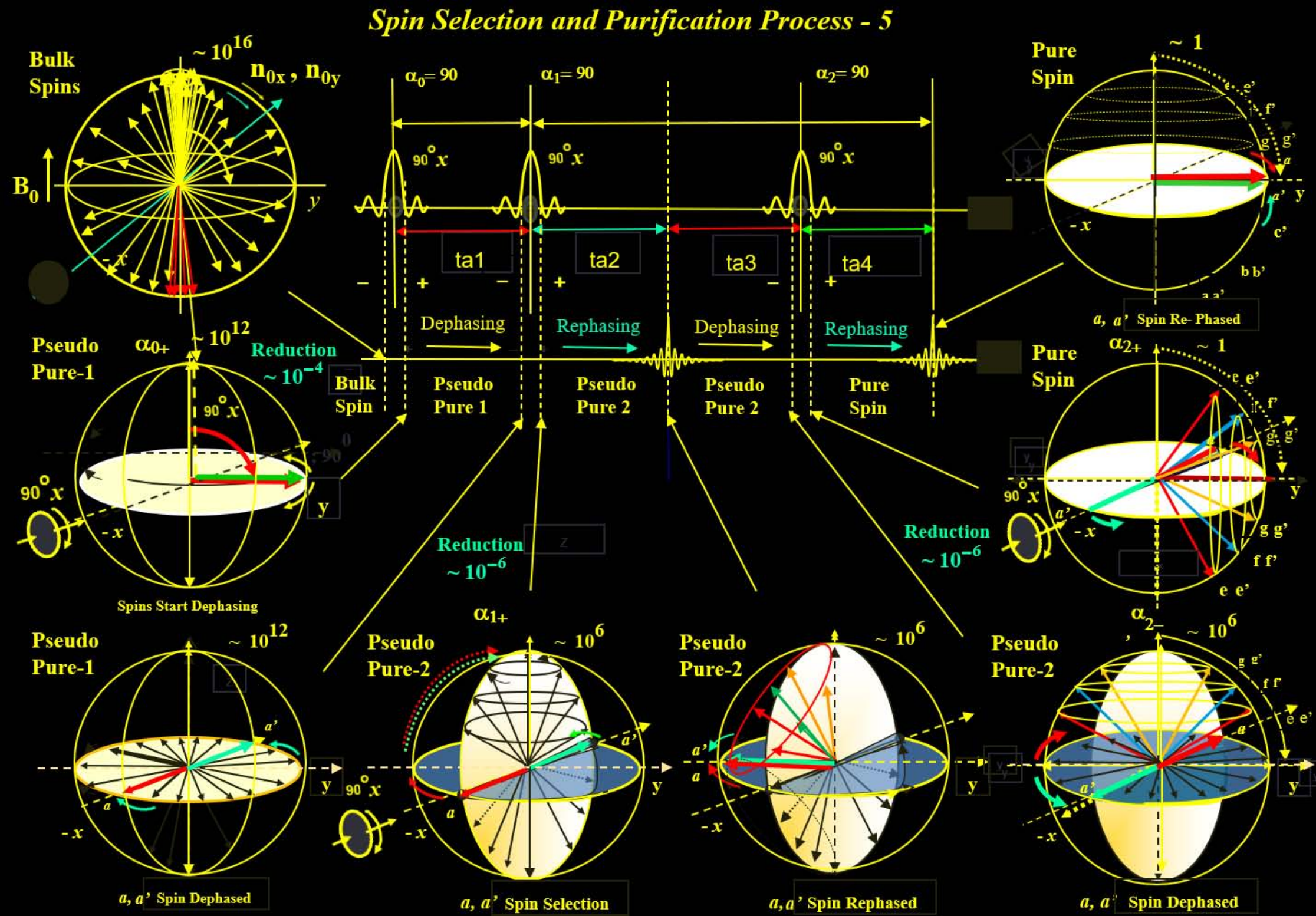}
  \caption{Purification technique using multiple $90^\circ$ rotations, known as the ``Stimulated Echo (STE)'' technique. Note the multiple $90^\circ$ rotations effectively select the a-a' spin pairs and eventually lead to the ``Pure'' state. Here, we assumed each purification factor as $10^{-6}$. To successfully achieve a high degree of purification, it is important to control the time intervals between the $90^\circ$ RF pulses, $t_{a1}, t_{a2}, \ldots$, as well as environmental noises that disturb the spin locations at the desired positions.}
\end{figure}

\begin{figure}[t]
  \centering
  \includegraphics[width=14cm]{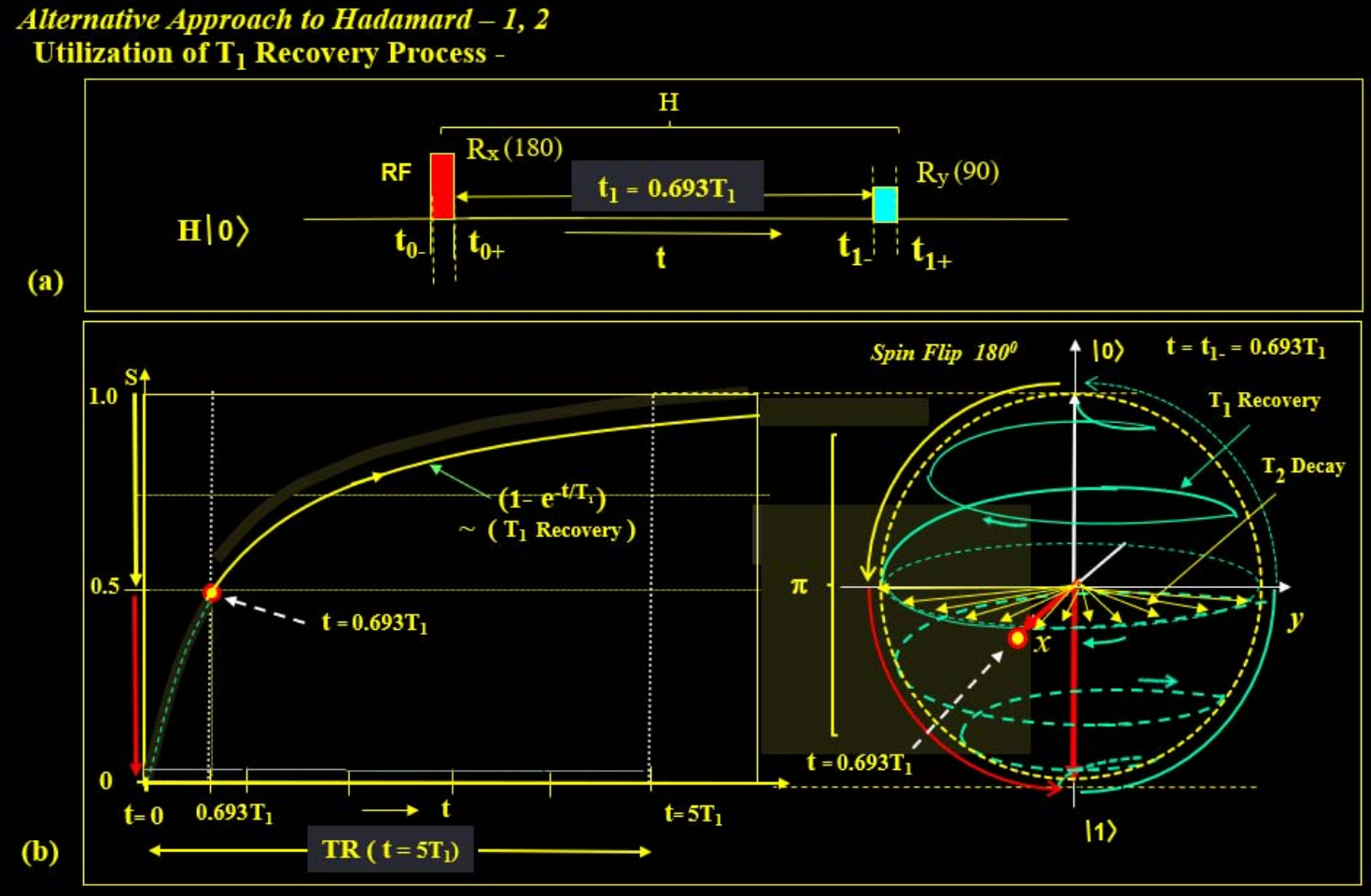}
  \caption{(a) Pulse sequence of the proposed alternative form of the Hadamard Gate using the T1 recovery process together with a simultaneously applied Time-Coincidence technique. \\
  (b) T1 recovery process and spin recovery trajectory in the Bloch sphere. Note the time, $t_1 = 0.693 T_1$, which we have used for the Time-Coincidence with the second RF, $R_y(90^\circ)$ for the $n_{0x}$ component.}
\end{figure}

\begin{figure}[t]
  \centering
  \includegraphics[width=14cm]{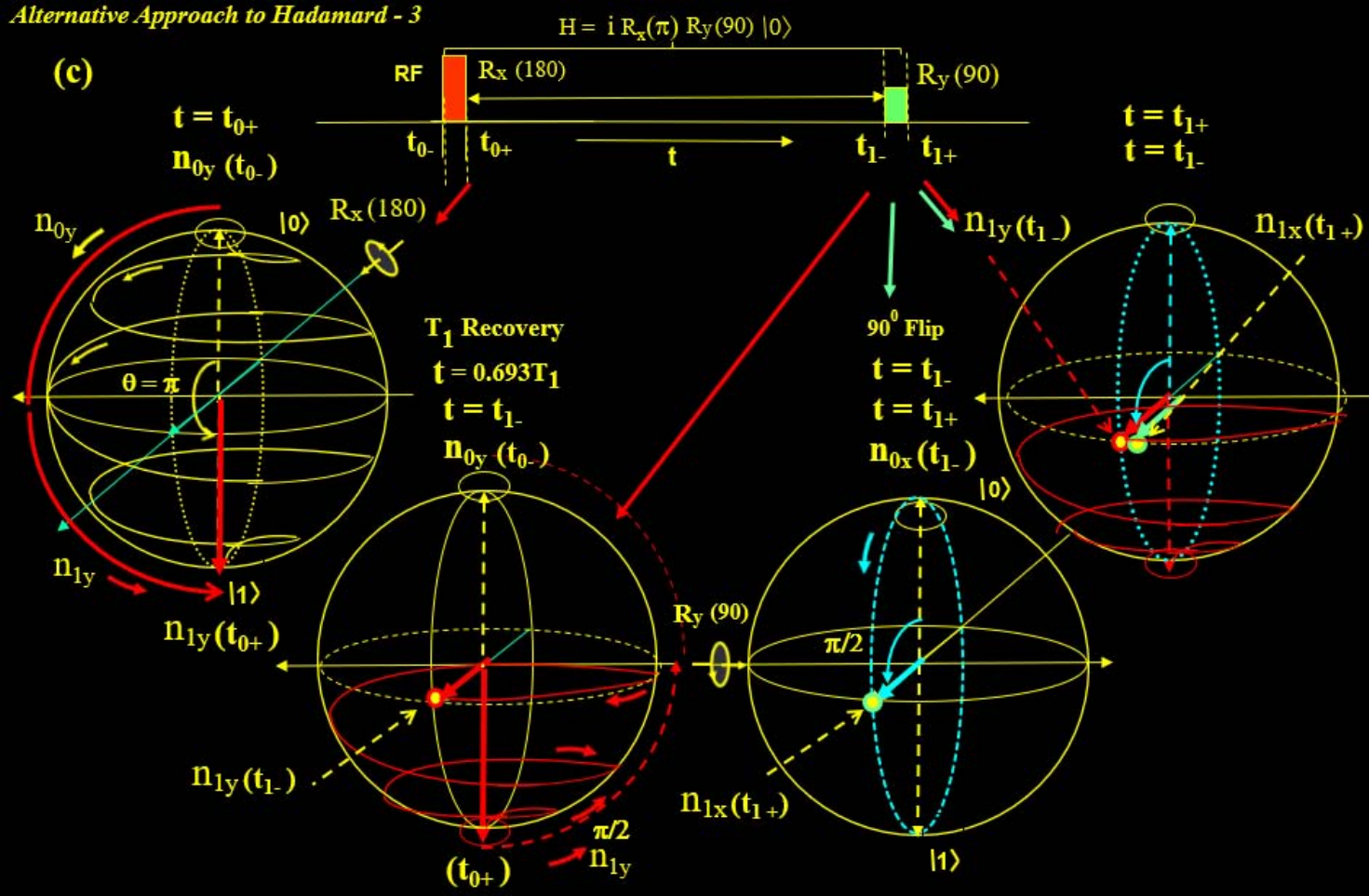}
  \caption{Pulse sequence used for the formation of the proposed alternative Hadamard Gate operation using T1 Recovery and Time-Coincidence. Note the resulting ``Superposition'' state at the end, $t = t_1^+$, that is, T1 Recovery process at $t_1 = 0.693 T_1$ with $R_x(180^\circ)$ and $R_y(90^\circ)$ RF pulses for the $n_{0x}$ component are in coincidence to form the Hadamard operation.}
\end{figure}

\begin{figure}[t]
  \centering

  \begin{subfigure}[t]{10cm}
    \centering
    \includegraphics[width=10cm]{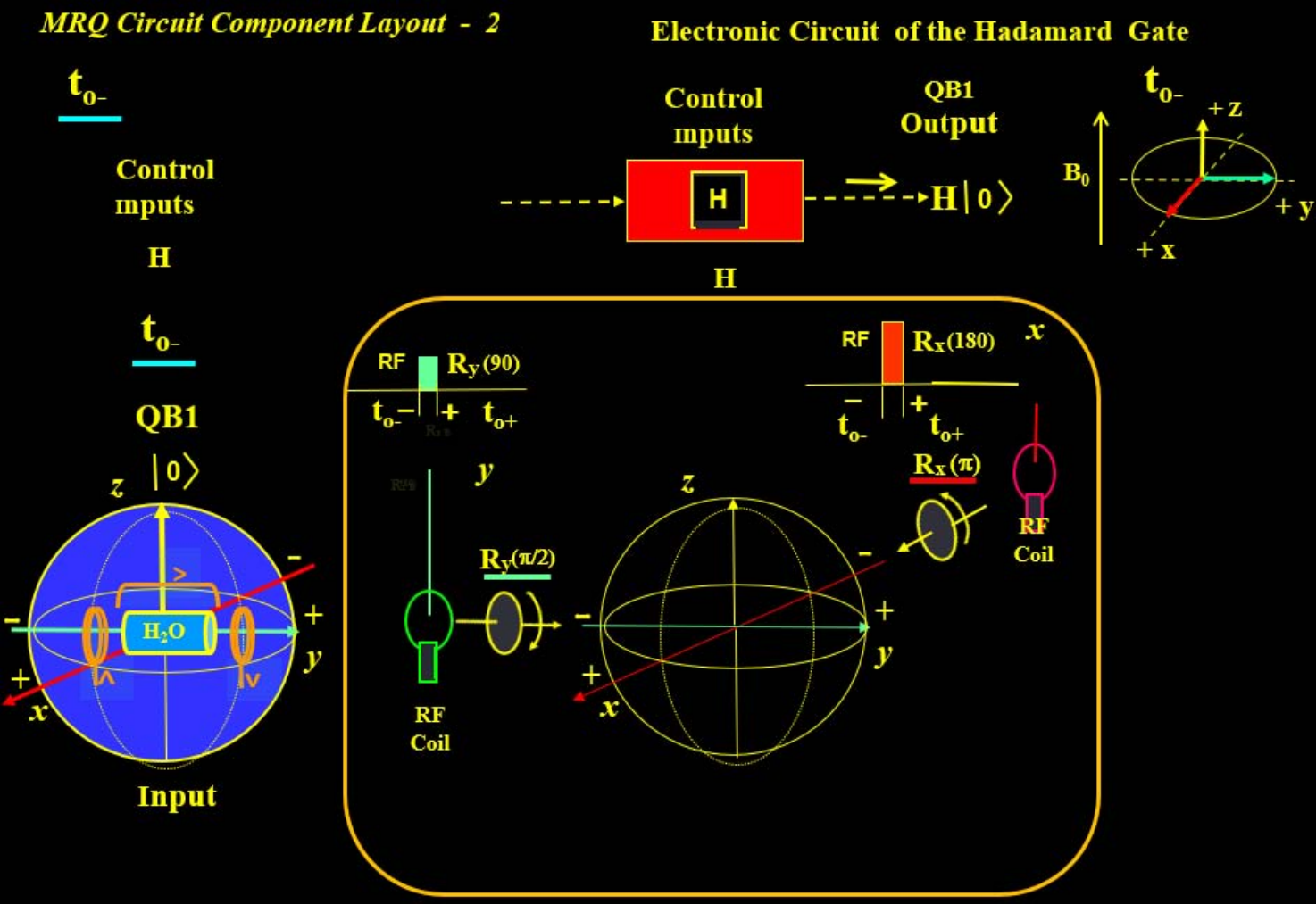}
    \caption{A Qubit and Input-Output Coupling \& Circuit Diagram for the Hadamard Gate Operation at initial state $t = t_0^-$.}
    \label{fig:App6}
  \end{subfigure}
  \vskip\baselineskip

  \begin{subfigure}[t]{10cm}
    \centering
    \includegraphics[width=10cm]{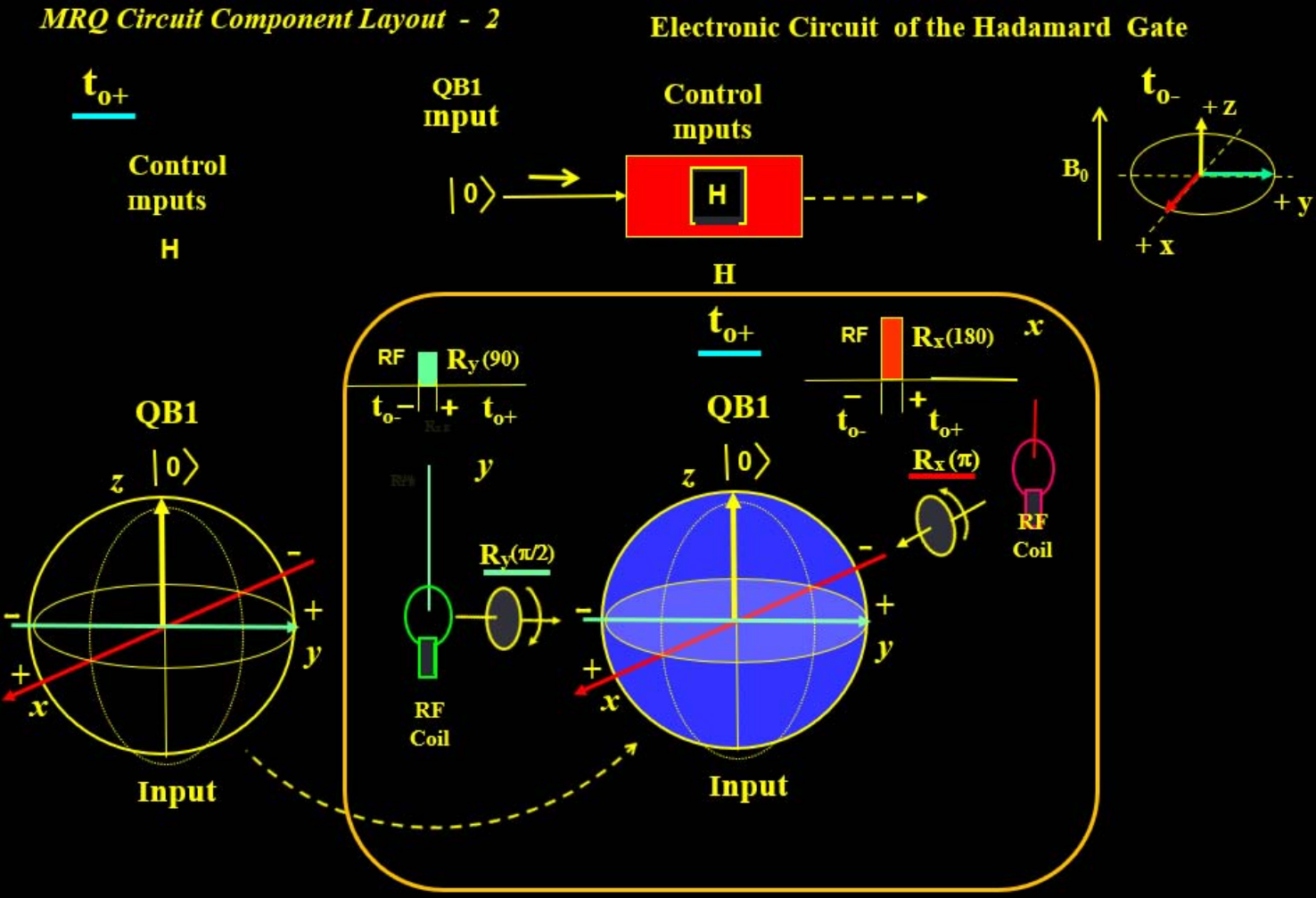}
    \caption{A Qubit and Input-Output Coupling \& Circuit Diagram for the Hadamard Gate Operation at time, $t = t_0^+$.}
    \label{fig:App6b}
  \end{subfigure}

 \begin{subfigure}[t]{10cm}
    \centering
    \includegraphics[width=10cm]{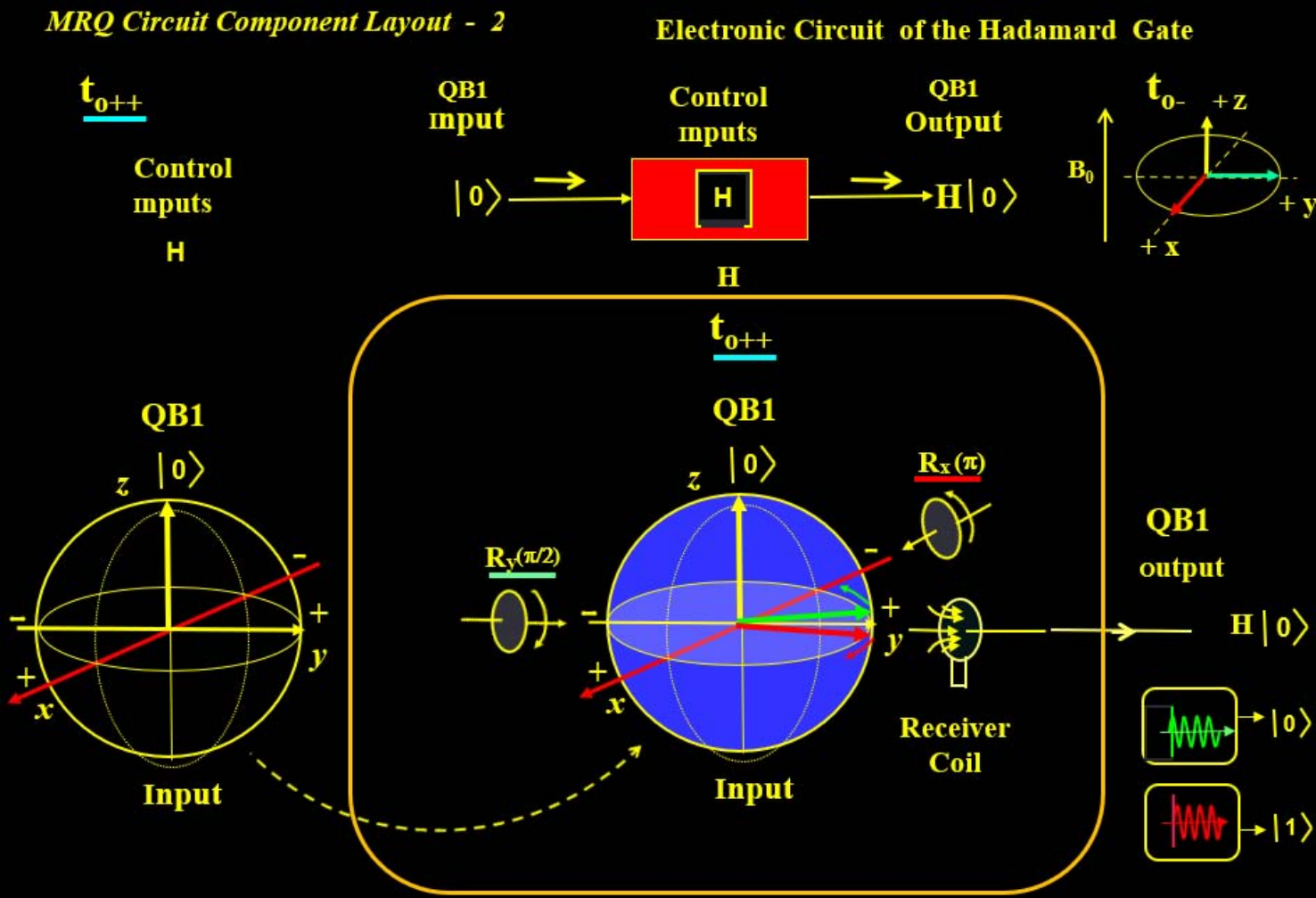}
    \caption{A Qubit and Input-Output Coupling \& Circuit Diagram for the Hadamard Gate Operation at time, $t = t_0^{++}$.}
    \label{fig:App6c}
  \end{subfigure}

  \caption{A Qubit and Input-Output Coupling \& Circuit Diagram for the Hadamard Gate Operation.}
  \label{fig:App6}
\end{figure}

\end{appendices}

\end{document}